\newcount\Comments  % 0 suppresses notes to selves in text
\documentclass[twocolumn]{aastex631}   

\usepackage{hyperref}
\usepackage{soul}
\usepackage{xcolor}
\usepackage{graphicx}
\usepackage{amsmath}
\usepackage{booktabs}
\usepackage{parskip}
\usepackage{bm}
\usepackage{makecell}

\newcommand{\boldell}{\bm{\ell}}
\newcommand{\boldL}{\bm{L}}

\definecolor{purple}{rgb}{0.6,0,1}
\definecolor{blue2}{rgb}{0,.5,1}
\definecolor{orange}{rgb}{1,0.6,0}
\newcommand{\kibitz}[2]{\ifnum\Comments=1\textcolor{#1}{#2}\fi}

\newcommand{\atau}{$180 \pm 157$}
\newcommand{\atauUL}{$450$}
\newcommand{\atauconsUL}{$370$}

\newcommand{\aCB}{$0.05 \pm 0.06$}
\newcommand{\aCBUL}{$0.15$}
\newcommand{\gUL}{$4.9 \times 10^{-2} / H_\mathrm{I}$}

\graphicspath{{./}{figures/}}

\newcommand{\DELETED}[1]{\relax}%                                               
{\relax}%

\shorttitle{$\tau$ and $\alpha$ constraints with ACT DR6} 
\shortauthors{Kramer et al.}

\begin{document}

\title{The Atacama Cosmology Telescope: \\Constraints on the anisotropic screening and birefringence effects with DR6}

\author[0000-0003-0238-8806]{Darby M. Kramer}
\affiliation{School of Earth and Space Exploration,
Arizona State University, 
Tempe, AZ 85287, USA}

\author[0000-0002-3495-158X]{Alexander van Engelen}
\affiliation{School of Earth and Space Exploration,
Arizona State University, 
Tempe, AZ 85287, USA}

\author[0000-0001-7805-1068]{Frank J. Qu}
\affiliation{Kavli Institute for Particle Astrophysics and Cosmology, Stanford University, 452 Lomita Mall, Stanford, CA 94305, USA}

\author[0000-0001-9420-7384]{Christopher Cain}
\affiliation{School of Earth and Space Exploration, Arizona State University, Tempe, AZ 85287, USA}

\author[0000-0003-3230-4589]{Irene Abril-Cabezas}
\affiliation{DAMTP, Centre for Mathematical Sciences, University of Cambridge, Wilberforce Road, Cambridge CB3 OWA, UK}
\affiliation{Kavli Institute for Cosmology Cambridge, Madingley Road, Cambridge CB3 0HA, UK}

\author[0000-0003-2358-9949]{J. Richard Bond}
\affiliation{CITA, University of Toronto, Toronto ON M5S 3H8 Canada}

\author[0000-0002-3589-8637]{Vera Gluscevic}
\affiliation{Department of Physics and Astronomy, University of Southern California, Los Angeles, CA 90089-0484}

\author[0000-0002-4765-3426]{Carlos Herv\'ias-Caimapo}
\affiliation{Instituto de Astrof\'isica and Centro de Astro-Ingenier\'ia, Facultad de F\'isica, Pontificia Universidad Cat\'olica de Chile, Av. Vicu\~na Mackenna 4860, 7820436 Macul, Santiago, Chile}

\author[0000-0002-8998-3909]{Niall MacCrann}
\affiliation{DAMTP, Centre for Mathematical Sciences, University of Cambridge, Wilberforce Road, Cambridge CB3 OWA, UK}
\affiliation{Kavli Institute for Cosmology Cambridge, Madingley Road, Cambridge CB3 0HA, UK}

\author[0000-0001-6740-5350]{Mathew S. Madhavacheril}
\affiliation{Department of Physics and Astronomy, University of Pennsylvania, 209 South 33rd Street, Philadelphia, PA 19104, USA}

\author[0000-0002-7245-4541]{Jeff McMahon}
\affiliation{Kavli Institute for Cosmological Physics, University of Chicago, 5640 South Ellis Avenue, Chicago, IL, 60637, USA}
\affiliation{Department of Astronomy and Astrophysics, University of Chicago, 5640 South Ellis Avenue, Chicago, IL, 60637, USA}
\affiliation{Department of Physics, University of Chicago, 5640 South Ellis Avenue, Chicago, IL, 60637, USA}

\author[0000-0003-3070-9240]{Toshiya Namikawa}
\affiliation{Center for Data-Driven Discovery, Kavli IPMU (WPI), UTIAS, The University of Tokyo, Kashiwa, 277-8583, Japan}

\author[0000-0001-6541-9265]{Bruce Partridge}
\affiliation{Department of Physics and Astronomy, Haverford College, Haverford, PA 19041, USA}

\author[0000-0002-4619-8927]{Emmanuel Schaan}
\affiliation{Kavli Institute for Particle Astrophysics and Cosmology, Stanford University, 452 Lomita Mall, Stanford, CA, 94305, USA}
\affiliation{SLAC National Accelerator Laboratory, 2575 Sand Hill Road, Menlo Park, California 94025, USA}

\author[0000-0002-9674-4527]{Neelima Sehgal}
\affiliation{Physics and Astronomy Department, Stony Brook University, Stony Brook, NY 11794}

\author[0000-0002-8149-1352]{Crist\'obal Sif\'on}
\affiliation{Instituto de F\'isica, Pontificia Universidad Cat\'olica de Valpara\'iso, Casilla 4059, Valpara\'iso, Chile}

\author[0000-0002-7567-4451]{Ed Wollack} \affiliation{Observational Cosmology Laboratory, NASA Goddard Space Flight Center, 8800 Greenbelt Road, Greenbelt, MD 20771}

\begin{abstract} 
While we have learned much from recent surveys of the mm-wave sky, many secondary anisotropies of the CMB remain under-explored. We consider two of these here: the anisotropic screening induced by Thomson scattering of CMB photons off free electrons, and cosmic birefringence, the rotation of CMB polarization due to parity-violating physics beyond the standard model. To measure these effects, we apply state-of-the-art estimators for statistical anisotropy to the ACT Data Release 6 CMB data and construct noisy maps and power spectra of both effects. For anisotropic screening, we find no significant detection with the baseline data, and place an upper limit on its power spectrum that is two orders of magnitude above predictions from current reionization models. We obtain evidence for some foreground contamination at one of our frequencies that is consistent with expectations from extragalactic simulations. For our anisotropic birefringence analysis, we also find no detection and place an upper limit on the signal that is consistent with previous analyses. We report a corresponding upper limit on the Chern-Simons coupling term $g_{a \gamma} <$ \gUL, with $H_\mathrm{I}$ the inflationary Hubble scale. These analyses pave the way toward obtaining improved constraints with the next generation of ground-based CMB surveys. 

\end{abstract}

\section{Introduction}\label{sec:intro}
Secondary anisotropies arise from interactions that the cosmic microwave background (CMB) photons have on their journey through the Universe to our telescopes, such as gravitational lensing, scattering with free electrons, and any other late-time interactions. Gravitational lensing of CMB photons by large scale structure has been extensively studied, leading to accurate measurements of its power spectrum and effect on the primordial CMB power spectrum (e.g., \textsl{Planck}, \citealt{plancklens2020}; ACT, \citealt{Qu2024}; SPT, \citealt{Ge2025}, as well as a joint measurement combining  all three \citealt{k5yr-3h6d}). Another type of secondary CMB anisotropy is the Sunyaev-Zel'dovich (SZ) effect, caused by inverse Compton scattering of CMB photons due to either thermal (tSZ) or bulk (kinetic or kSZ) velocities in electron halos that surround galaxies and clusters. These effects have been measured and detected in various ways many times \citep[for a recent review, see][]{Mroczkowski2019}. There are many secondary effects that have not yet been detected, including two particular secondaries of current interest: anisotropic screening and birefringence. 

\paragraph{Anisotropic Screening}
Thomson scattering, the low-energy limit of Compton scattering, occurs when photons interact with free electrons. Specifically, the electrons absorb photons, accelerate, and re-emit photons at the same wavelength but along different lines of sight, which washes out the incident CMB anisotropy. Unlike CMB lensing and the SZ effects, the anisotropic ``screening'' of the CMB caused by free electrons has not been robustly detected. Because there are multiple sources of free electrons in the Universe, this effect is important in both cosmology and astrophysics.

One of these sources is the epoch of reionization (EoR), a period of time in the Universe's history when the first stars and galaxies were able to photo-ionize the intergalactic medium (IGM), creating bubbles of free electrons corresponding to scales of $\sim 10-100$ comoving Mpc \citep[e.g.,][]{lin2016}. The morphology and timing of the EoR are uncertain, though it is likely that it ended by redshift 5--6, according to Lyman alpha observations \citep[e.g.,][]{lyalpha2006,redshift6,Bosman2022}. This reionization process is believed to have been inhomogeneous, or ``patchy,'' instead of uniform due to the discrete objects that fueled it \citep[e.g., ][]{inhomoLoeb2001}. Because bubbles of free electrons interact with photons via Thomson scattering, the CMB is expected to contain a ``patchy screening'' signal, among others (such as the patchy kSZ signal), from the EoR \citep{dvorkin2009,gluscevic2012}. 

The shape and amplitude of the patchy screening power spectrum depends heavily on both the morphology and timing of reionization \citep[e.g.,][]{dvorkinsmith2009}. The dependency is similar to how the patchy kSZ power spectrum constrains reionization; measurements of the total kSZ power spectrum from small-scale CMB surveys have placed appreciable limits on the duration of reionization \citep{Reichardt_2021, beringue2025, chaubal2026}. This makes both the kSZ and screening spectra highly sought-after reionization probes. The screening power spectrum amplitude is particularly sensitive to the midpoint and duration of reionization due to its line of sight integral over the reionization history \citep{kramer2025}. Its shape is indicative of a characteristic angular bubble scale, to which the bubbles of free electrons grew before coalescing into a fully-ionized universe \citep[e.g.,][]{furlanetto2006}. Therefore, constraining this power spectrum can yield critical new information on the evolution of reionization. 

The post-EoR Universe also contributes to the anisotropic screening signal because massive galaxies retain halos of free electrons. The CMB scattering from the ionized gas in low-redshift halos has been detected via the kSZ effect \citep[e.g.,][and many more]{Hand2012,SPTksz2016,qu2026,hadzhiyska2026}, but the radial profiles of these halos are still not well-quantified due to their extended, low-density nature. Because the properties of the electron halos depend strongly on unconstrained baryonic feedback processes, it is important to push these measurements further. Both low-redshift galaxy halos and reionization bubbles contribute to the total anisotropic screening signal, with the patchy contribution from reionization expected to be dominant in the power spectrum by $\sim$ 1 order of magnitude \citep[e.g.,][]{roy2022probing}. There is a strong measurement of the spatially-averaged value of the electron optical depth, most recently of $\bar{\tau} = 0.058 \pm 0.006$ with the \textsl{Planck} satellite \citep{Tristram2024}. However, the anisotropy  of $\tau$ is as-yet undetected.

The power spectrum of the screening effect can be reconstructed with CMB data using quadratic estimators in a very similar manner to CMB lensing \citep{dvorkinsmith2009, suyadav2011}. Constraints on the screening power spectrum have been made with previous experiments including \textsl{WMAP} \citep{gluscevic2012}, \textsl{Planck} \citep{toshiyatau2018}, and BICEP/Keck \citep{Ade2023}. A recent search for post-EoR anisotropic screening in cross-correlation with unWISE galaxies found no significant detection \citep{Coulton2025}, and no experiments to-date have the sensitivity to measure the screening auto-power spectrum based on current preferred models. The upcoming generation of CMB experiments are forecasted to measure this signal at 3--4$\sigma$, or up to 20$\sigma$ depending on the true reionization history \citep{dvorkinsmith2009, roy2018, Bianchini2023,Jain2024}. Still, placing upper limits on the amplitude of screening informs on allowed ranges of parameters in models for both reionization and baryonic feedback.
 
\paragraph{Anisotropic Cosmic Birefringence} Cosmic birefringence is a rotation of the linear CMB polarization that would arise from certain parity-violating extensions of the standard model coupling to photons, converting their $E$ modes into $B$ modes and vice versa. It could manifest as a uniform (isotropic) rotation angle across the sky, as spatially-varying (anisotropic) fluctuations in the rotation field, or as a combination of both, depending on the underlying mechanism. The isotropic component is accessible through two-point functions of the CMB: using the CMB $EB$ cross-power spectrum, \citet{minamikombirefnonzero} measured a hint of a nonzero mean rotation angle of $0.35 \pm 0.14$~deg with \textsl{Planck} data, a result that has persisted in some subsequent analyses \citep[e.g.,][]{diegopalazuelos2022,EskiltandKomatsu_2022,Eskilt2023Cosmoglobe,diegopalazuelos2025}. The anisotropic component, by  contrast, induces off-diagonal CMB mode-coupling and is reconstructed via quadratic estimators in a manner analogous to CMB lensing. No detection of anisotropic birefringence has been made to-date. Aside from earlier studies \citep[e.g.,][]{polarbear2017,bicep_biref2017, contreras2017}, current 95\% upper limits on the amplitude of a scale-invariant spectrum (defined in Equation~\ref{eq:acbdef} and discussed in Section~\ref{sec:bireftheory} below) include $A_\mathrm{CB} < 0.1$ from ACT~DR4 \citep{namikawa2020} and SPTpol \citep{Bianchini_2020}, $A_\mathrm{CB} < 0.044$ from BICEP/Keck \citep{Ade2023}. Various \textsl{Planck} analyses have also placed competitive constraints on anisotropic birefringence \citep{Gruppuso_2020,Bortolami_2022,zagatti2024}, with the tightest upper limit reported by \citep{Bortolami_2022} at $A_\mathrm{CB} < 0.021$. Searching for this signal in new data and improving these upper limits are worthwhile efforts in constraining the theoretical models that could produce birefringence, particularly in light of the persistent hint of a non-zero isotropic signal.

\paragraph{This paper} In this work, we build dedicated quadratic estimators for anisotropic screening and cosmic birefringence within the framework of the ACT DR6 lensing pipeline developed in \citet{Qu2024}. We then apply these to this dataset. The high angular resolution and small-scale sensitivity of ACT make it particularly well-suited to measure new sources of off-diagonal mode-coupling in the CMB beyond lensing, and to place strong constraints on these signals. Previous studies have utilized the ACT DR6 data for CMB lensing \citep{Qu2024} and other secondary signals \citep{maccrann2024}; here we extend this program to screening and birefringence. We generate maps of both effects; estimate their power spectra; and after subtracting well-understood sources of power spectrum bias, we compare our resulting power spectrum band-powers with theoretical models. This work contains the first constraints on the power spectrum of anisotropic screening using the small-scale CMB fluctuations that are accessible from ground-based surveys. Because of the use of these small scales, we also perform a detailed screening foreground analysis with mock data.

We describe the data and simulations used in this work in Sections~\ref{sec:data} and \ref{sec:sims}, respectively. Section~\ref{sec:theory} contains the screening (\ref{sec:tautheory}) and birefringence (\ref{sec:bireftheory}) theory. In Section~\ref{sec:methods}, we summarize the derivation of a general quadratic estimator of the CMB in the context of screening and cosmic birefringence (\ref{subsec:recon}) before discussing details of the cross-split estimator we use here (\ref{subsec:crossest}). The subsections~\ref{sec:mapbiases} and \ref{sec:psbiases} discuss each bias that we calculate and remove from the screening and birefringence reconstructions, including those induced by CMB lensing, which are large biases with respect to the signals of interest. We describe the pipeline implementation and verification in Section~\ref{sec:implement}. Section~\ref{sec:results} contains the band-power constraints for screening and birefringence with ACT DR6. Section~\ref{sec:systematics} details our studies of systematics and consistency tests in each analysis. Section~\ref{sec:disc} provides interpretations of our analyses and discusses the outlook for screening and birefringence measurements. Finally, we summarize our study, results, and conclusions in Section~\ref{sec:conc}.

\section{Data}\label{sec:data}
The Atacama Cosmology Telescope (ACT) was a 6 meter ground-based telescope located in the Atacama Desert of Chile \citep{2007ACTspecs, ACTpolspecs2016}. It surveyed the CMB temperature and polarization across almost half of the sky at multiple frequencies with high resolution, gathering smaller scale information than space-based CMB surveys.\footnote{See the \hyperlink{https://lambda.gsfc.nasa.gov/product/expt/}{LAMBDA - CMB Experiments} page for info on many past, present, future, and canceled CMB experiments}

ACT Data Release 6 (DR6) is the last major data release from ACT \citep{naess2025, louis2025, calabrese2025}. This data release consists of maps that cover $\sim 45\%$ of the sky with 1.4 arcminute resolution, and to depths that are roughly a factor of 2 better than the previous ACT DR5 \citep{dr5_2020}. Our analysis benefits strongly from the previous DR6 lensing analysis, \citet[][hereafter Q24]{Qu2024}, which completed several data processing steps beyond the extensive map-making from \citet{naess2025}. We use the same map products throughout for consistency with the Q24 simulation suite and bias calculations.\footnote{We use the first science-grade version of the ACT DR6 maps, labeled \texttt{dr6.01}. A refined version of the maps, \texttt{dr6.02}, which improves the large-scale transfer function and polarization noise levels and includes data taken in 2022, has since been produced and is used for the DR6 public CMB data release. We do not use \texttt{dr6.02} here because our bias calculations and covariance estimates rely on the extensive set of simulations built to match the \texttt{dr6.01} products.} Specifically, Q24 combines data from separate arrays into 90 and 150 GHz splits, deconvolves the beams in each set, corrects for the transfer function induced by detector gain variations, and self-calibrates the polarization amplitude. The final pre-processing step includes subtracting point-sources using ACT DR5 + \textsl{Planck} data, and subtracting tSZ clusters and additional sources found using the source-finding package, NEMO.\footnote{\href{https://nemo-sz.readthedocs.io/}{https://nemo-sz.readthedocs.io/}}

Next, as part of the DR6 lensing pipeline, Q24 masked the data to remove dominant contaminants. First, a real space mask was applied to remove regions with strong Galactic contamination. Here, we adopt the \textsl{Planck} 60\% mask for our baseline case as done in Q24 (we test the robustness of this choice in Section \ref{sec:const_tests}). This map was then masked to select the ACT DR6 footprint, which was defined by removing regions of the data where the noise is $> 70\mu \mathrm{K}^\prime$ for either 90 or 150 GHz, and removing any leftover strong Galactic contaminants such as dust clouds. This left a sky fraction of approximately 23\%, or 9400 deg$^2$, for science analysis. Finally, the mask was apodized by a cosine taper with a 3 degree width. Separately, a Fourier-space mask was applied to remove Fourier modes $-90 < \ell_x < 90$ and $-50 < \ell_y < 50$, which are contaminated by ground pickup; this is a standard practice in ACT analyses \citep[see, e.g., ][]{louis2017}. Since the maps were stored with 0.5 arcminute pixels, Q24 downgraded them to 1 arcminute resolution and window-deconvolved them. Some extended or bright sources remained, so the maps  were masked and inpainted at these locations. Finally, they were coadded by frequency in preparation for analysis. See Q24 Sections 3 and 5 for details about the extensive preparation of the data for this work.

\section{Simulations}\label{sec:sims}
For most of our bias calculations (Sections~\ref{sec:mapbiases} and \ref{sec:psbiases}), we use the same simulation suite as Q24. These simulations contain CMB signal, lensing, Gaussian extragalactic foreground power, and noise realizations whose spatially-varying properties are derived from the data to match the statistics of the ACT survey and scan strategy \citep{atkins2023}. They are frequency-coadded and Fourier-space filtered in the same manner as the data. Crucially, while they contain the CMB lensing effect, these simulations do not contain anisotropic screening or birefringence signals. This is by design: the dominant biases to our reconstructions arise from the mask, CMB, and lensing, all of which are present in these simulations. Excluding the target signal is in fact necessary for the additive Monte Carlo correction of Section~\ref{sec:psbiases}, whose expectation value is zero only for signal-free simulations.

In order to perform tests and calculate biases that require the desired signal to be present in simulations, such as the mask-induced transfer function laid out in Section~\ref{sec:tf}, we additionally construct a set of 100 noiseless simulations that have both lensed CMB and a simulated Gaussian random field of anisotropic screening, as well as a separate set with lensed CMB and a Gaussian anisotropic rotation field applied. Since these are both as-yet-undetected signals, we use theory-based power-spectra to generate the Gaussian random fields. For anisotropic screening, we use a smoothed version of the \citet{battaglia2013} simulation's patchy screening power spectrum from the reionization era\footnote{Note that the EoR-based patchy screening is expected to dominate the total anisotropic screening power spectrum, including that from lower redshifts.}. For the birefringence simulations, we use a scale-invariant power spectrum, as defined below in Equation \ref{eq:acbdef} below, with an amplitude\footnote{This amplitude represents 95\% the upper limit obtained with ACT DR4 \citep{namikawa2020}.} of $A_\mathrm{CB} = 0.1$. Both sets are Fourier-space filtered in the same way as the data. This is necessary because the filtering suppresses power in the reconstructed signals; we use these simulations to measure the transfer function (Section~\ref{sec:tf}), to then correct for this suppression in the final band-powers.

\section{Theory}\label{sec:theory}
\begin{figure*}[htp!]
\hspace{-15pt}
\includegraphics[width=\textwidth]{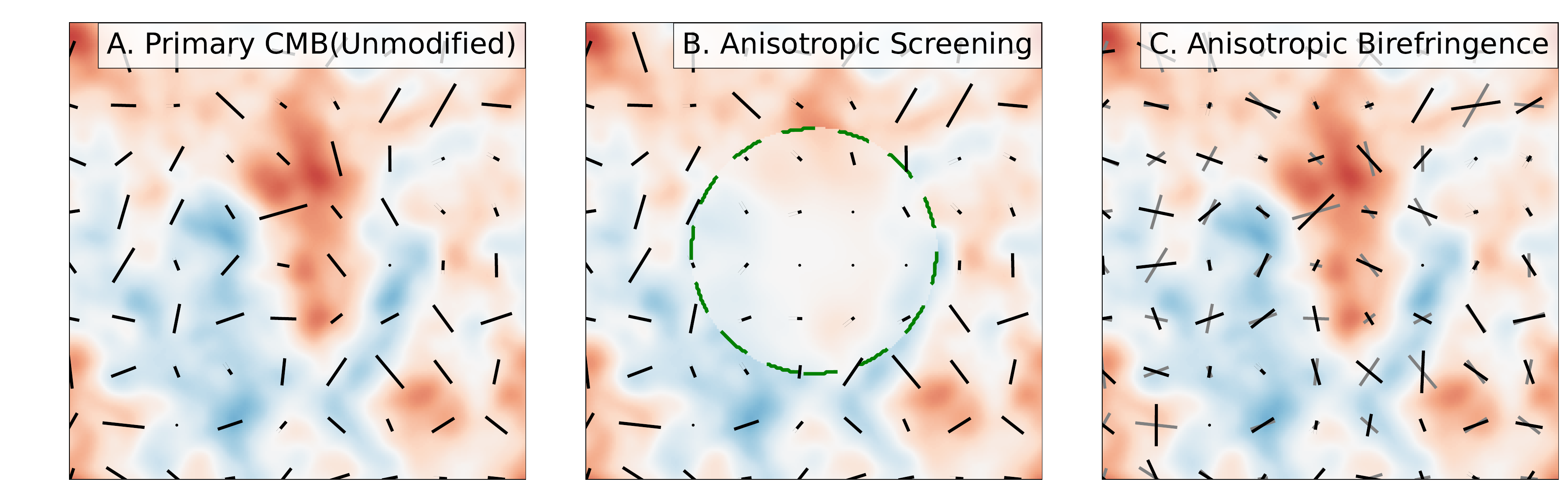}
    \caption{Demonstration of exaggerated map-level effects of anisotropic screening and birefringence on the CMB. Panel A contains the primary, unmodified CMB temperature (red and blue) and polarization (black headless vectors) from a simulation. Panel B is what happens to the primary CMB if a dense blob of electrons is placed within the green dashed circle; the temperature and polarization signal along those lines of sight gets washed out. Panel C shows the effect if instead there is a Gaussian random anisotropic rotation field applied to the CMB; the new gray vectors show that each original polarization vector has incurred a line-of-sight-dependent rotation angle.}\label{fig:taudemo}
\end{figure*}

\subsection{Anisotropic Screening Theory}\label{sec:tautheory}
The optical depth of electrons in a given direction in the Universe can be written as the line-of-sight integral of the average electron number density $\bar{n}_e$ and the electron density fluctuations $\delta_e(\mathbf{r}, \eta)$ at comoving position $\mathrm{r}$ and conformal time $\eta$:
\begin{equation}\label{eq:tau_n}
    \tau(\hat{\mathbf{n}}) = c\ \sigma_T \int_0^{\eta_{\mathrm{CMB}}} d\eta\ a(\eta)\ \bar{n}_e(\eta) \left[1 + \delta_e(c\eta \hat{\mathbf{n}},\eta_0 -\eta)\right],
\end{equation}
where $\eta$ is conformal lookback time, $\sigma_T$ is the Thomson scattering cross section, $\bar{n}_e(\eta)$ is the spatially-averaged electron number density at a given $\eta$, and $a(\eta)$ is the scale factor.

This anisotropic optical depth affects the observed CMB temperature signal as the modulation $\tau(\mathbf{\hat{n}})$:
\begin{equation}
    T(\mathbf{\hat{n}}) = e^{- \tau(\mathbf{\hat{n}})} \tilde{T}(\mathbf{\hat{n}}), \label{eq:tttaumod}
\end{equation}
and the $Q$ and $U$ Stokes polarization parameters as
\begin{equation}
    (Q \pm iU) (\mathbf{\hat{n}}) = e^{- \tau(\mathbf{\hat{n}})}\ (\tilde{Q} \pm i\tilde{U}) (\mathbf{\hat{n}}) \label{eq:qutaumod},
\end{equation}
where a tilde represents the observed CMB fields in the absence of optical depth modulation. Screening manifests as washed-out temperature and polarization fluctuations in CMB maps, as demonstrated in Panel B of Figure~\ref{fig:taudemo}.

\subsection{Cosmic Birefringence Theory}\label{sec:bireftheory}
Cosmic birefringence could be caused by primordial magnetic fields (PMFs) or pseudo-scalar fields (PSFs) that may exist or have existed in the Universe which would interact with photons and change their polarization over time \citep{kamionkowski2009}. 

PMFs would have caused Faraday rotation, which arises from the differing refractive indices of the two circular polarization states in a magnetized plasma and therefore scales as $\nu^{-2}$, yielding anisotropic rotation that varies between frequency channels \citep{yadav2012}. Rotation from a PSF, by contrast, arises from the modification to the photon dispersion relation itself rather than from the intervening medium, and is achromatic. The spatial structure is a separate question: if the PSF is uniformly-distributed throughout the Universe like dark energy, its effect on CMB photons would be isotropic. However, if the PSF interacts with matter through, e.g., gravity, or is generated in a manner that leads to fluctuations in its energy density, it would be an anisotropic effect and likely would trace the matter overdensities. These distinctions make the possible causes separable in observations with enough sensitivity. In this work, we implement methods that constrain the anisotropic effect from PSFs only, such that we can combine the 90 and 150 GHz DR6 data. We do not constrain PMFs in this work for simplicity, as it would require separate frequency analyses.

If linear CMB polarization rotation exists and is frequency-independent but anisotropic, it could lead to the discovery of the PSF as an axion-like field with mass $10^{-33} \leq m_a \leq 10^{-28}$ eV \citep[for a review of this topic, see][]{axcosm2016}. In this case, the PSF couples with electromagnetic (EM) fields, and thus photons, through a Chern-Simons term, $g_{a\gamma}$, on the EM tensor strength \citep[e.g., ][]{carroll1998}:
    \begin{equation}
        \mathcal{L} \supset \frac{g_{a\gamma}a}{4} F_{\mu\nu}\tilde{F}^{\mu \nu},
    \end{equation}
    
    where $ a $ represents the axion-like particle field. The rotation angle is then defined in terms of the change in the field $\Delta a$ over a given line-of-sight:
    \begin{equation}
        \alpha = \frac{g_{a \gamma}}{2} \Delta a.
    \end{equation}
    In the limit where the axion was massless during inflation, its field would be represented by a scale invariant power spectrum and can be written in the limit of large ($L \lesssim 100$) reconstructed scales as \citep{caldwell2011}:
    \begin{equation}
        \frac{L(L+1)C_L^{\alpha \alpha}}{2 \pi} = \left ( \frac{H_\mathrm{I} g_{a \gamma}}{4 \pi} \right )^2
        \label{eq:claadef}
    \end{equation}
    with $H_\mathrm{I}$ as the inflationary Hubble parameter. We choose to represent this power spectrum with a dimensionless amplitude, $A_\mathrm{CB}$, as done in \citet{namikawa2020}:
    \begin{equation}
    \begin{aligned}
        C_L^{\alpha \alpha} = & A_\mathrm{CB} \frac{2\pi}{L(L+1)} \times 10^{-4}\ [\mathrm{rad}^2]. \label{eq:acbdef}
    \end{aligned}
    \end{equation}
Cosmic birefringence is the rotation of the CMB polarization fields in a given direction by an angle $\alpha$ in the following way:
\begin{equation}
    (Q \pm iU) (\mathbf{\hat{n}}) = e^{\pm 2i\alpha(\mathbf{\hat{n}})}\ (\tilde{Q} \pm i\tilde{U}) (\mathbf{\hat{n}}),\label{eq:qualphamod}
\end{equation}
which therefore rotates CMB E-mode polarization into B-modes, and vice versa. We constrain the anisotropic component of this effect in this work, which is demonstrated in Panel C of Figure~\ref{fig:taudemo}. Note the similarities in Equations \ref{eq:qutaumod} (screening) and \ref{eq:qualphamod} (birefringence): the difference is only in the factor of $-\tau$ versus $2i\alpha$ in the pre-factor. This mathematical similarity motivates measuring both effects together with the same pipeline in a single analysis.

\section{Methods}\label{sec:methods}
In this section, we present our screening and birefringence reconstruction methods. The analysis in this paper is performed on the curved sky using spherical harmonic transforms. For notational brevity, however, we present the estimators and their components in their flat sky approximations. The transformations from real space CMB temperature and polarization maps $[T,\ Q,\ U]$ to their Fourier counterparts are:
\begin{multline}\label{eq:cmbfourier}
    T(\boldell) = \int T(\hat{\mathbf n})\ e^{-i\boldell \cdot \hat{\mathbf n}}\ d\hat{\mathbf n};\\
    Q(\boldell) = \int Q(\hat{\mathbf n})\ e^{-i\boldell \cdot \hat{\mathbf n}}\ d\hat{\mathbf n};\\
    \mathrm{and}\ U(\boldell) = \int U(\mathbf{\hat{n}})\ e^{-i\boldell \cdot \hat{\mathbf n}}\ d\hat{\mathbf n}
\end{multline}
and the transformations from $[Q,\ U]$ to $[E,\ B]$ are:
\begin{equation}\label{eq:qutoeb}
    [E(\boldell)\pm iB(\boldell)] = [Q(\boldell)\pm iU(\boldell)]\ e^{\mp 2 i \varphi_{\boldell}}\ 
\end{equation}
where $\varphi_{\boldsymbol{\ell}}$ is the polar angle of the $\boldsymbol{\ell}$ vector in the 2D Fourier plane (i.e.\ the angle between $\boldsymbol{\ell}$ and the $\ell_x$ axis). \citet{dvorkinsmith2009} and \citet{kamionkowski2009} derive the more general curved sky versions of these equations for anisotropic screening and cosmic birefringence, respectively. Using the {\tt{tempura}}\footnote{\href{https://github.com/simonsobs/tempura}{https://github.com/simonsobs/tempura}} and {\tt{falafel}}\footnote{\href{https://github.com/simonsobs/falafel}{https://github.com/simonsobs/falafel}} pipelines, which perform general quadratic estimators of the CMB (including lensing), we expand the larger Q24 lensing pipeline to include the screening and birefringence estimators we describe below. This enables simultaneous reconstruction of lensing alongside either signal. 

\subsection{General Quadratic Estimator Reconstruction}\label{subsec:recon}

Here we introduce a generic field $\Gamma$ that generates statistical anisotropy and produces off-diagonal mode-coupling of the CMB. Anisotropic screening, cosmic birefringence, and CMB lensing cause off-diagonal correlations between CMB modes $\ell_1$ and $\ell_2$ in the following way:
\begin{equation}
     \langle X(\boldell_1) Y(\boldell_2)\rangle_\mathrm{CMB}\ = f^{\Gamma}_{XY}(\boldell_1,\boldell_2) \ \Gamma(\boldL),
 \end{equation}
with a fixed field $\Gamma$ and mode-coupling $f^{\Gamma}_{XY}(\boldell_1,\boldell_2)$ given in Table \ref{tab:filters} for screening, birefringence, and lensing. Each $X$ and $Y$ can be replaced with a $T$, $E$, or $B$. One can reconstruct this  $\Gamma$ field in Fourier space using the quadratic estimator first derived in \citet{huokamoto2002}. This estimator has been adopted for other CMB secondaries since its derivation, in e.g. \citet{dvorkinsmith2009, kamionkowski2009}. We adopt the notation for the equivalent flat sky approximation presented in \citet{suyadav2011}, where an optimal estimator for a $\Gamma$ field is given by:
\begin{equation}
    \begin{split}
    & \hat{\Gamma}_{XY}(\boldL) = \\
    & A_{XY}(\boldL) \int \frac{d^2 \boldell_1}{(2\pi)^2} [X(\boldell_1)Y(\boldell_2)]F^{\Gamma}_{XY}(\boldell_1,\boldell_2).
    \end{split} \label{eq:taurecon}
\end{equation}
In this work, $\boldL = \boldell_1 + \boldell_2$ (or more importantly, $\boldell_2 = \boldL - \boldell_1$, where $\boldL$ is held fixed over the integrals of $\boldell_1$ and $\boldell_2$). $A_{XY}(\boldL)$ is the normalization function that ensures the estimate is unbiased from the filtering function $F^{\Gamma}_{XY}(\boldell_1,\boldell_2)$: 
\begin{equation} 
    A_{XY}^{\Gamma}(\boldL) = \left[ \int \frac{d^2 \boldell_1}{(2\pi)^2} F^{\Gamma}_{XY}(\boldell_1,\boldell_2) f^{\Gamma}_{XY}(\boldell_1,\boldell_2) \right]^{-1} \label{eq:taunorm},
\end{equation}
\begin{equation}
    F^{\Gamma}_{XY}(\boldell_1,\boldell_2) = \frac{f^{\Gamma}_{XY}(\boldell_1,\boldell_2)}{2C_{\ell_1}^{XX\mathrm{,tot}} C_{\ell_2}^{YY\mathrm{,tot}}} \label{eq:taucapf}.
\end{equation}

where $C_{\ell}^{XX\mathrm{,tot}}$ is the sum of the CMB, noise, and foreground power spectra $C_{\ell}^{XX\mathrm{,tot}} = C_{\ell}^{XX} + C_{\ell}^{XX, \text{noise}} + C_{\ell}^{XX, \text{foregrounds}}$. For CMB fields $X$ and $Y$ where $X \neq Y$, the $2$ in the denominator is not present \citep{huokamoto2002}. The mode-coupling expressions, $f^{\Gamma}_{XY}(\boldell_1,\boldell_2)$, which contain the prescription for the CMB modes that the field $\Gamma$ couples, are given in Table \ref{tab:filters} for anisotropic screening, cosmic birefringence, and CMB lensing for all possible combinations of mode couplings. As a first check of the analysis pipeline, we verified that it exactly reproduces the Q24 lensing band-powers from both simulations and data, while enabling these new reconstructions.

\begin{figure}
    \includegraphics[width=1.05\linewidth]{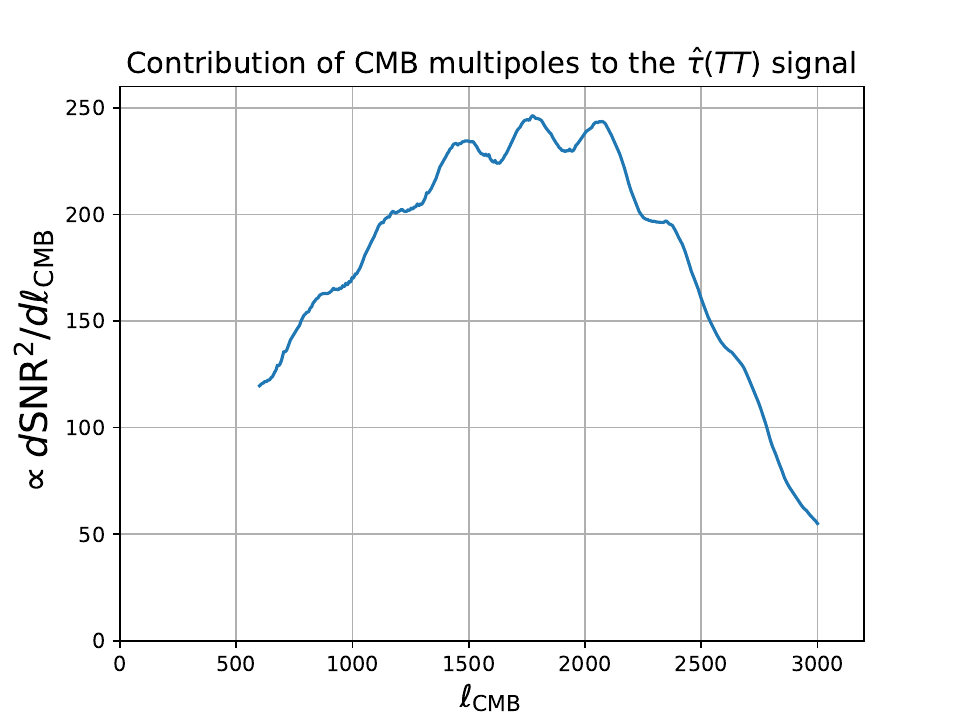}
    \includegraphics[width=1.05\linewidth]{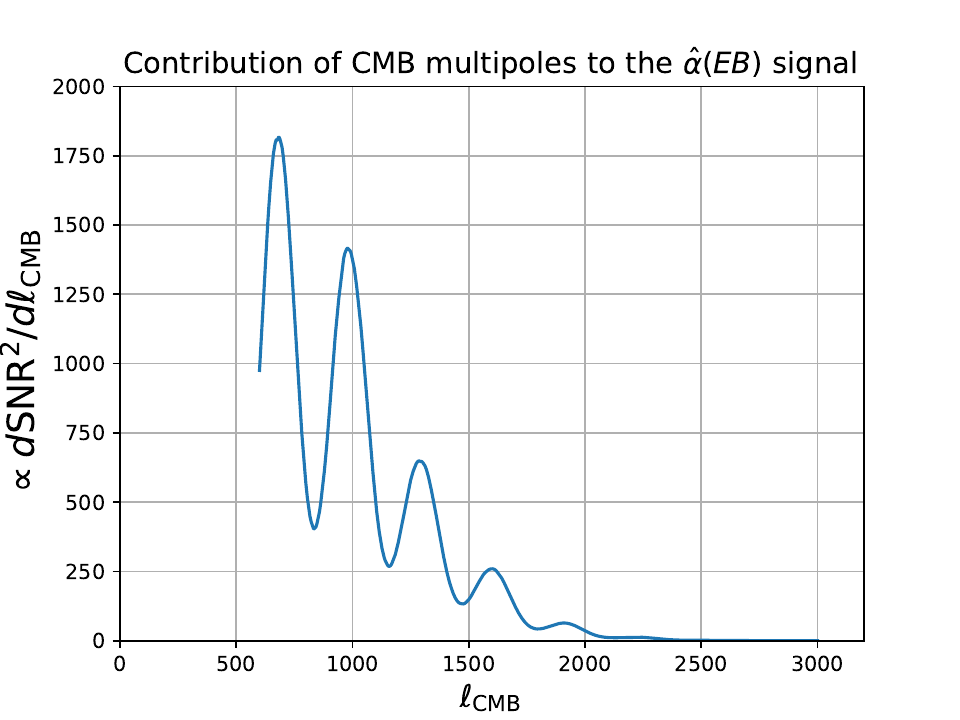}
    \caption{Demonstration of the sensitivity of the screening (\emph{top}) and cosmic birefringence (\emph{bottom}) estimators to different CMB modes under the squeezed limit, i.e. for large-scale $\tau$ and $\alpha$ fluctuations. This plot is the integrand of Equation~\ref{eq:taunorm}, and represents the contribution of each CMB multipole to the total SNR$^2$ per reconstructed mode for screening and birefringence. We have truncated these curves below $\ell_\mathrm{CMB} = 600$ and above $\ell_\mathrm{CMB} = 3000$ because those are the $\ell_\mathrm{min}$ and $\ell_\mathrm{max}$ chosen in this analysis. In both cases, measuring large-scale CMB modes well is helpful, especially for birefringence estimators. However, ACT DR6 has strong sensitivity at the important scales for anisotropic screening measurements.}
    \label{fig:filters}
\end{figure}

We choose to use only the CMB $TT$ quadratic estimator for the anisotropic screening analysis due to the relatively-large polarization noise level in the ACT DR6 data. In the future, e.g. with data from the Simons Observatory \citep[SO,][]{soscience2019}, better CMB polarization sensitivity will be essential to improving the SNR on anisotropic screening measurements. For the birefringence analysis, we use the $EB$ estimator since birefringence is a polarization effect.

For the filters applied in our reconstructions, we set the total power spectra $C_{\ell}^{XX\mathrm{,tot}}$ to match the total power in the DR6 data, identical to the choices in Q24. We base our CMB multipole ranges on several factors. The ACT DR6 power spectrum analysis, \citet{louis2025}, chose their minimum CMB multipole to be $\ell_\mathrm{min} = 600$, which we follow in our reconstructions. We choose $\ell_\mathrm{max} = 3000$ because foregrounds become dominant over the CMB power at scales smaller than this. 

To demonstrate the CMB Fourier modes that are most important for the screening and birefringence estimators, in Figure~\ref{fig:filters} we plot the integrand of Equation \ref{eq:taunorm} under the squeezed limit, i.e. that $\ell_1 \approxeq \ell_2$ and $\ell_1, \ell_2 >> L$. This effectively estimates the contribution of each CMB multipole to the total signal-to-noise per mode of the reconstructed map. For screening, this indicates that the DR6 reconstruction is sensitive to intermediate CMB scales and that we do not lose much constraining power in our choice of $\ell_\mathrm{min} = 600$. However, for cosmic birefringence we see that the highest sensitivity is at the largest CMB scales, so we are strongly limited by our $\ell_\mathrm{min}$ restriction.

\begin{table*}[htp!]
    \centering
    \begin{tabular}{c|c|c|c}
         \hline
         $XY$ &
         $f^{\tau}_{XY}(\boldsymbol{\ell}_1,\boldsymbol{\ell}_2)$ &
         $f^{\alpha}_{XY}(\boldsymbol{\ell}_1,\boldsymbol{\ell}_2)$ &
         $f^{\phi}_{XY}(\boldsymbol{\ell}_1,\boldsymbol{\ell}_2)$
         \\[0.5ex]
         \hline\hline
         $TT$
         & $C_{\ell_1}^{TT}+C_{\ell_2}^{TT}$
         & ---
         & $(\boldsymbol{L}\!\cdot\!\boldsymbol{\ell}_1)C_{\ell_1}^{TT}+(\boldsymbol{L}\!\cdot\!\boldsymbol{\ell}_2)C_{\ell_2}^{TT}$
         \\
         \hline
         $EB$
         & $[C_{\ell_1}^{EE}-C_{\ell_2}^{BB}]\sin\!\big[2\varphi_{\boldsymbol{\ell}_1\boldsymbol{\ell}_2}\big]$
         & $2\,C_{\ell_1}^{EE}\cos\!\big[2\varphi_{\boldsymbol{\ell}_1\boldsymbol{\ell}_2}\big]$
         & $[(\boldsymbol{L}\!\cdot\!\boldsymbol{\ell}_1)C_{\ell_1}^{EE}-(\boldsymbol{L}\!\cdot\!\boldsymbol{\ell}_2)C_{\ell_2}^{BB}]\sin\!\big[2\varphi_{\boldsymbol{\ell}_1\boldsymbol{\ell}_2}\big]$
         \\
         \hline
    \end{tabular}
    \caption{
    Minimum-variance mode-couplings for screening ($\tau$), cosmic birefringence ($\alpha$), and CMB lensing ($\phi$).
    Here $\varphi_{\ell_1} = \arctan(\ell_{1 y}/\ell_{1 x})$ is the polar angle of the vector $\boldell_1$ in the Fourier plane, $\varphi_{\ell_1 \ell_2} \equiv \varphi_{\ell_1} - \varphi_{\ell_2}$, and $\boldL = \boldell_1+\boldell_2 $.}
    \label{tab:filters}
\end{table*}

\subsection{Cross-Split Estimator}\label{subsec:crossest}
The standard quadratic estimator applied to a single coadded map produces a power spectrum that is biased by the noise properties of the data, requiring accurate noise simulations to debias the result. As demonstrated in Q24, this is difficult to achieve for ground-based data with complex, spatially-varying atmospheric noise; any mis-modelling of the noise propagates directly into the bias subtraction and can contaminate the result. To avoid this, we use the cross-split estimator introduced in \citet{madhavacheril2020}, which is robust to noise mis-modelling by construction. After coaddition of eight data splits into four, we begin the analysis with the data $T$, $E$, and $B$ each divided into four splits with independent noise realizations. Following \citet{madhavacheril2020} and subsequently Q24, the quadratic estimator described in Section~\ref{subsec:recon} is applied to all non-redundant combinations of different splits and averaged. This operation ensures robustness against noise mis-modelling, and we summarize the details here. 

We reconstruct $\hat{\Gamma}$ from two CMB fields $X$ and $Y$, each available in four independent splits. For a given pair of distinct splits $(\mathrm{a},\mathrm{b})$ we symmetrize over the two leg assignments,
\begin{equation}
    \hat{\Gamma}_{XY}^{\rm{a},\rm{b}}(\boldL) = \frac{1}{2}\left[\hat{\Gamma}_{X^{\mathrm{a}}Y^{\mathrm{b}}}(\boldL) + \hat{\Gamma}_{X^{\mathrm{b}}Y^{\mathrm{a}}}(\boldL)\right],
    \label{eq:splitmaprecon_example}
\end{equation}
and the reconstructed map averages over the six split pairs that never reuse a split,
\begin{equation}
    \hat{\Gamma}_{XY}(\boldL) = \frac{1}{6}\sum_{a<b}\hat{\Gamma}_{XY}^{\rm{a},\rm{b}}(\boldL).
    \label{eq:splitmaprecon_full}
\end{equation}
We hereafter suppress the $XY$ subscript, taking $XY = TT$ for screening ($\tau$) and $XY = EB$ for birefringence ($\alpha$). The power spectrum is then estimated by averaging cross-spectra over every set of four distinct splits, so that no split is repeated within a single spectrum. Because only maps with independent instrument and atmospheric noise enter, the estimator is insensitive to noise mis-modelling in the mean-field and Gaussian bias terms:
\begin{equation}
    C_L^{\hat{\Gamma} \hat{\Gamma},\times}(XY, UV) = \frac{1}{4!} \sum_{a \neq b \neq c \neq d} \bar{C}_L^{\times}\!\left[\hat{\Gamma}_{XY}^{\rm{a},\rm{b}}(\boldL),\ \hat{\Gamma}_{UV}^{\rm{c},\rm{d}}(\boldL)\right],
    \label{eq:splitpsrecon_def}
\end{equation}
where $\bar{C}_L^{\times}[\hat{I}(\boldL), \hat{J}(\boldL)]$ denotes the cross-power spectrum of maps $\hat{I}(\boldL)$ and $\hat{J}(\boldL)$. Equation~\eqref{eq:splitpsrecon_def} is schematic: it indicates the split pairings retained, namely those in which no split is repeated across the four legs, but in practice we evaluate it using the accelerated form of \citep{madhavacheril2020}, which reorganizes the same sum into auxiliary estimators built from different split combinations. We drop $(XY, UV)$ below with the understanding that $(XY, UV) = (TT, TT)$ for screening and $(EB, EB)$ for birefringence. This entire analysis, other than where noted, was performed with this ``cross-split'' estimator.

\subsection{Map-level Biases}\label{sec:mapbiases}
There are two map-level biases introduced from the reconstruction of $\hat{\Gamma}(\boldL)$: the mean-field and the transfer function.

\subsubsection{Transfer Function}\label{sec:tf}
The masks applied during pre-processing (Section~\ref{sec:data}) remove CMB modes that would otherwise contribute to the reconstruction, suppressing the recovered signal. We characterize this suppression using the signal-injected, noise-free simulations of Section~\ref{sec:sims}, defining the map-level transfer function, $\mathrm{TF}_\mathrm{map}$, as the ratio of the cross-spectrum between the input signal (as a Gaussian random field) and its reconstruction, to the input signal's autospectrum:
\begin{equation}
    \mathrm{TF}_\mathrm{map} = 
    \frac{C_L^{\Gamma\hat{\Gamma}}}{C_L^{\Gamma\Gamma}}.
    \label{eq:tf_maplevel}
\end{equation}
Since the power spectrum is constructed from the square of the map, the corresponding power spectrum-level transfer function is the square of this quantity:
\begin{equation}
    \mathrm{TF} = \left( 
    \frac{C_L^{\Gamma\hat{\Gamma}}}{C_L^{\Gamma\Gamma}} 
    \right)^2.
    \label{eq:tf}
\end{equation}
In this analysis, the bias-subtracted band-powers and their error bars are divided by the binned transfer function, as indicated by the $1/\mathrm{TF}$ in Equation~\ref{eq:sum_recon} below. We show the map-level transfer function in Figure~\ref{fig:tfs}; the suppression is roughly 15--30\% at the map-level across all reconstructed scales. This is similar to the equivalent result for the case of CMB lensing found in Q24.

\begin{figure}
    \centering
    \includegraphics[width=\linewidth]{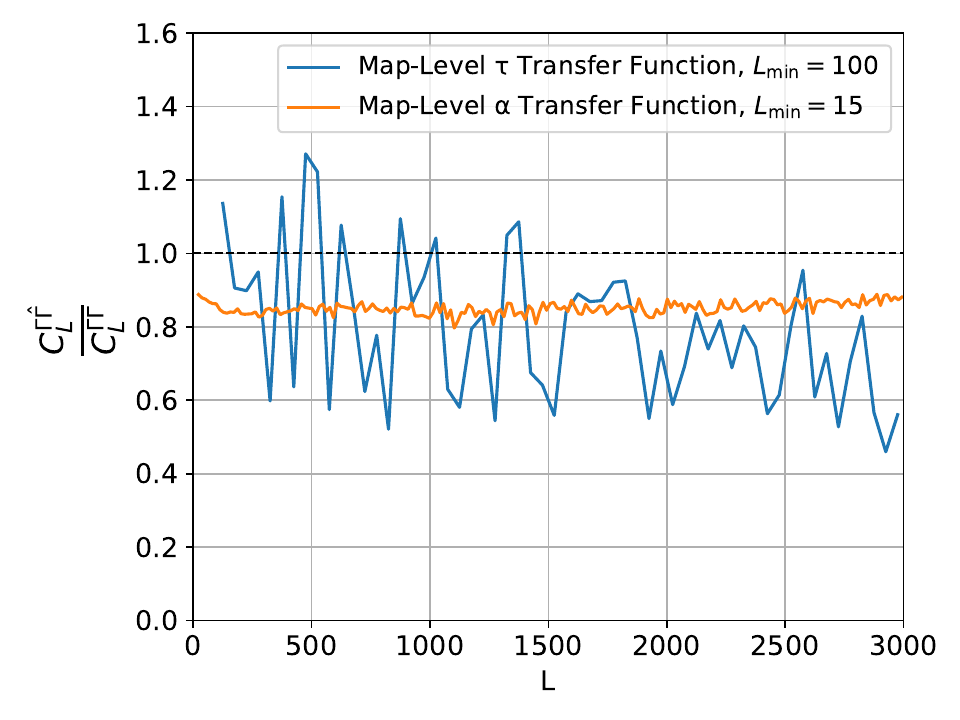}
    \caption{The binned map-level $\tau$ and $\alpha$ transfer functions defined in Equation~\ref{eq:tf_maplevel}. These represent the loss of power due to the Fourier-space filtering of the CMB modes (discussed in Section~\ref{sec:data}) before simulations and data enter the reconstruction pipeline, which is roughly a 15-30\% effect on all scales at the map level.}
    \label{fig:tfs}
\end{figure}

\subsubsection{Mean-field Subtraction}\label{subsec:mfbias}

\begin{figure}
    \includegraphics[width=0.49\textwidth,trim={3.7cm, 0, 2.5cm, 0}, clip]{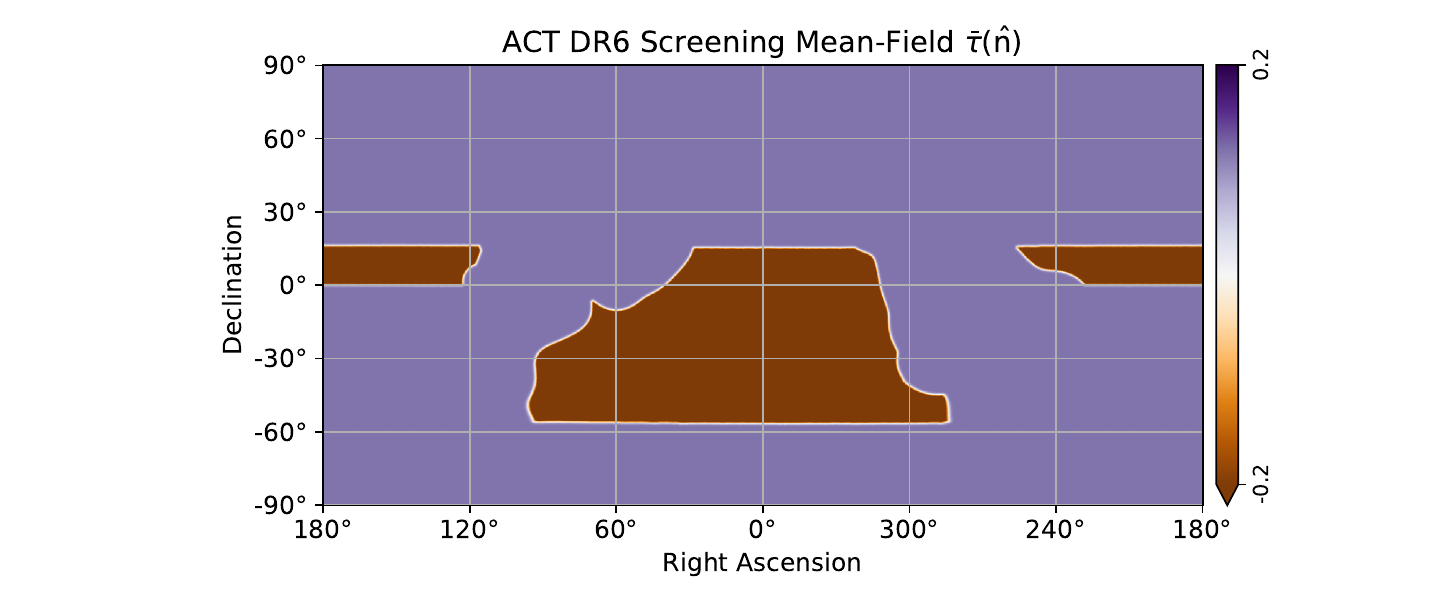}
    \caption{The reconstructed and normalized screening mean-field. This mean-field has large amplitudes, and mostly has fluctuations on the largest scales, which is consistent with the power spectrum of this map in Figure~\ref{fig:tau_biases}. This map has been smoothed with a 30 arcminute Gaussian kernel for visual purposes. Note that we do not show the equivalent mean-field for birefringence because its amplitude is sub-dominant (see the birefringence power spectrum-level biases in Figure~\ref{fig:rot_biases}).}
    \label{fig:mfmaps}
\end{figure}

\begin{figure*}
    \centering
    \includegraphics[width=\textwidth]{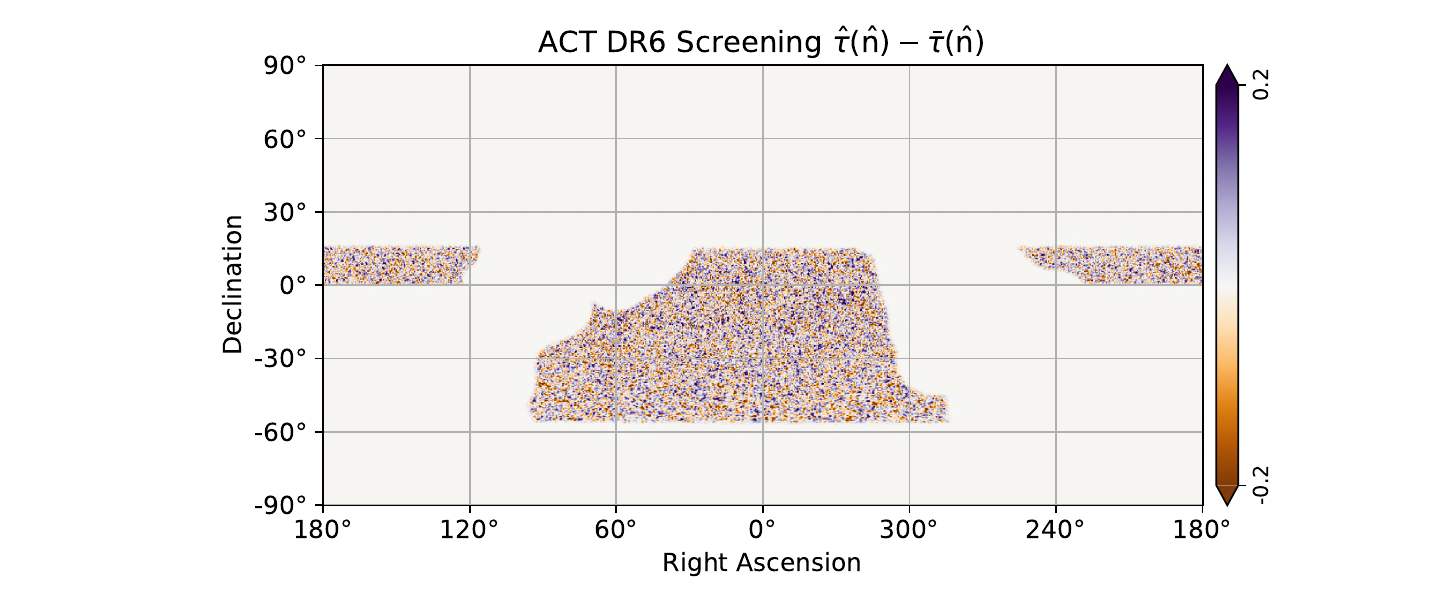}
    \caption{Reconstructed, normalized, mean-field subtracted ACT DR6 anisotropic screening map. This map is dominated by reconstruction noise at these sensitivities. It has been smoothed by a 30 arcminute Gaussian kernel for visual purposes.}
    \label{fig:taumap}
\end{figure*}

\begin{figure*}
    \includegraphics[width=\textwidth]{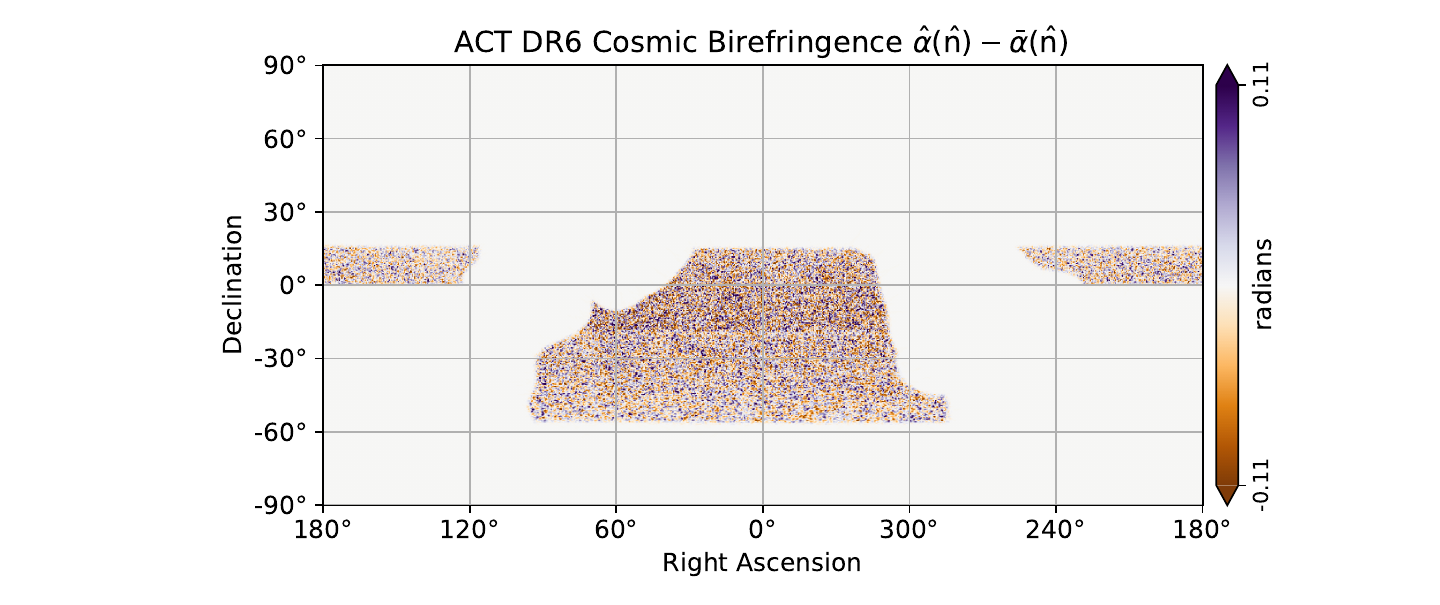}
    \caption{Reconstructed, normalized, mean-field subtracted ACT DR6 anisotropic birefringence map. This map, like the screening map, is dominated by reconstruction noise at these sensitivities. The feature at $\sim -20$ deg is due to the variation in depth of the initial CMB maps, particularly the B-mode map due to its lack of signal. See \citep{naess2025} for details on the varying depths of the DR6 data. This map has been smoothed by a 30 arcminute Gaussian kernel for visual purposes.}
    \label{fig:birefmap}
\end{figure*}

The real-space mask applied to the data (Section~\ref{sec:data}) breaks statistical isotropy, which the quadratic estimator picks up as a spurious signal. This ``mean-field bias'' must be estimated and subtracted at map-level, prior to taking the power spectrum (as depicted in Equation \ref{eq: mfsub_ps}). We estimate the mean-field by passing the Q24 simulation suite (Section~\ref{sec:sims}) through the same reconstruction pipeline as the data. The mean-field is then the average reconstruction over all simulations:
\begin{equation}
 \bar{\Gamma}(\boldL) = \frac{1}{N_\mathrm{sim}} 
    \sum_{i=1}^{N_\mathrm{sim}} 
    \hat{\Gamma}^{i}(\boldL).
\end{equation}
Each simulation has an independent CMB and lensing realization, and the cross-split estimator ensures that the noise also does not contribute. What remains is the spurious signal induced by the mask, which is common to all simulations. We use $N_\mathrm{sim} = 390$ simulations for this estimate. The resulting mean-field map for screening is shown in Figure~\ref{fig:mfmaps}. We choose not to show the corresponding mean-field map for birefringence because its features are invisible on the same color bar as the residual reconstructed map. See the birefringence power spectrum-level biases in Figure~\ref{fig:rot_biases} for a reference to how small this bias is. 

\begin{figure}
    \centering
    \includegraphics[width=\linewidth]{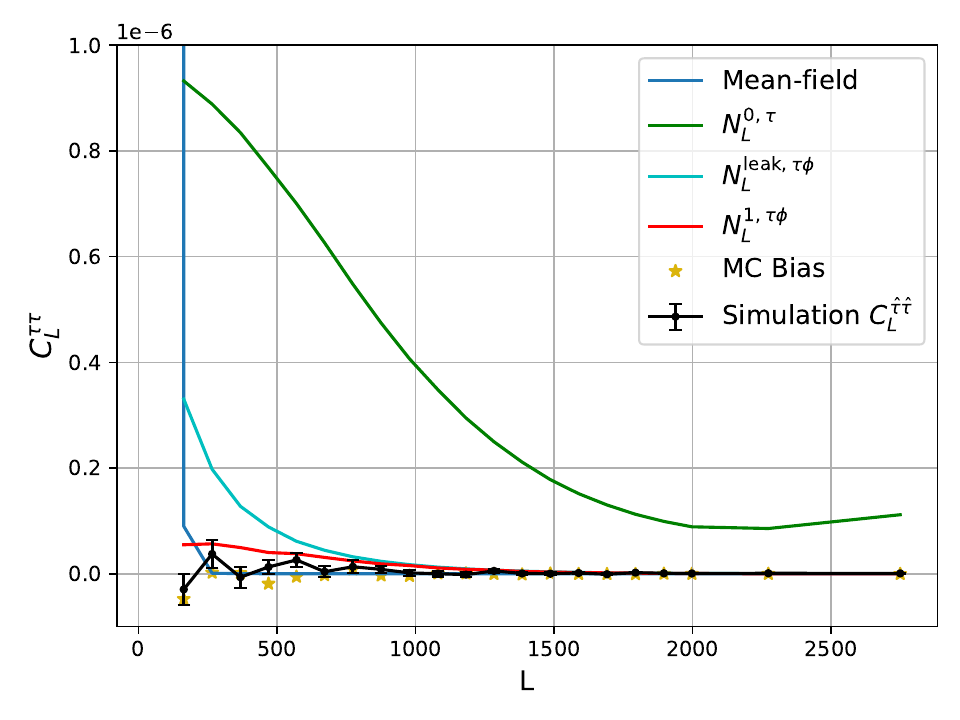}
    \caption{Power spectra of all the biases that are calculated and removed from $C_L^{\hat{\tau} \hat{\tau}}$. The largest bias at low-$L$ is the mean-field bias (dark blue), which arises from the real-space galaxy mask and noise anisotropy that induces mode-coupling via the quadratic estimator. The largest bias on all other scales is the $N_L^{0,\tau}$ or disconnected bias (green) due to random-chance couplings from the mostly-Gaussian CMB entering the quadratic estimator. The cyan bias is the power spectrum of the lensing bias, $N_L^{\mathrm{leak,}\tau\phi}$, that leaks into the $\tau$ estimator. The smaller $N_L^{1,\tau\phi}$ bias is shown in red, which represents the contamination from the secondary lensing trispectrum into the screening estimator. The gold stars represent the additive MC bias, or the remaining bias to the simulation reconstructions that is not explicitly quantified in this analysis. The bias-subtracted band-powers from one screening-free simulation are shown in black, and demonstrate that even though there are several large biases to the screening estimator, they are quantifiable and can be completely removed to yield an unbiased result that is consistent with zero. All biases are described in Sections \ref{sec:mapbiases} and \ref{sec:psbiases}.}
    \label{fig:tau_biases}
\end{figure}

\begin{figure}
    \centering
    \includegraphics[width=\linewidth]{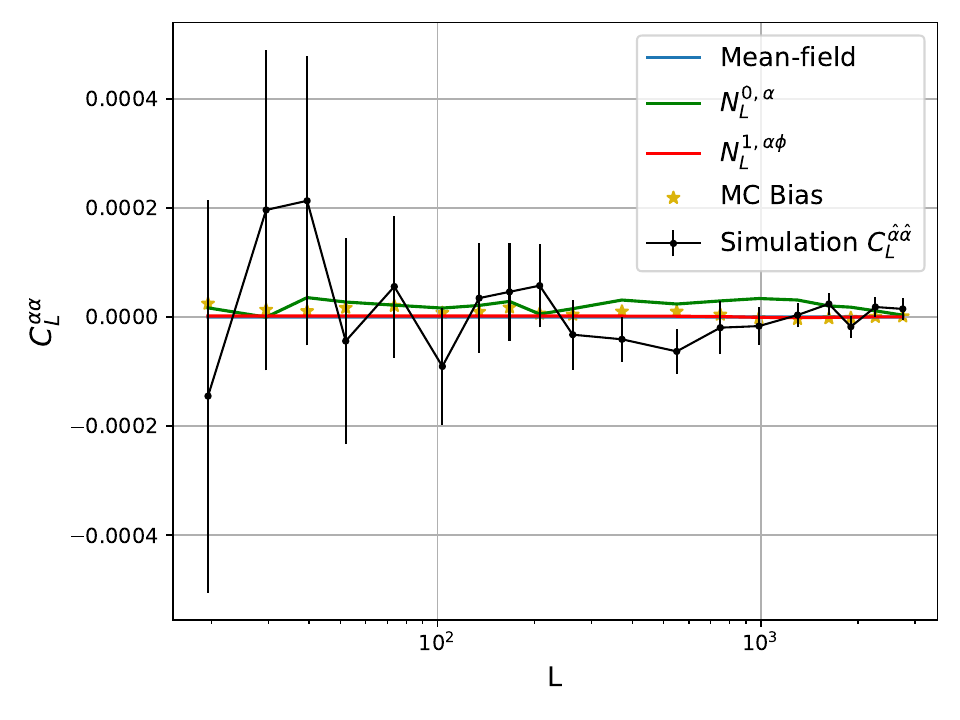}
    \caption{The same biases described in Figure~\ref{fig:tau_biases} but for anisotropic cosmic birefringence reconstruction (with a $log(L)$ x-axis to show the logarithmic bins more clearly). These error bars are much larger than the corresponding biases, in stark contrast to the screening errors, because the DR6 CMB polarization noise is much higher than that for temperature. The birefringence $N_L^{0,\alpha}$ bias is the most dominant, and is generally flatter than that for screening. We also note that the mean-field bias is negligible because the global polarization angle of the simulations is zero for these simulations. We separately address and handle this systematic bias in Section \ref{subsec:rot_mfbias}. }
    \label{fig:rot_biases}
\end{figure}

The ACT DR6 reconstructed, mean-field-subtracted screening map $\hat{\tau}(\boldL) - \bar{\tau}(\boldL)$ is shown in Figure~\ref{fig:taumap} and the corresponding birefringence map $\hat{\alpha}(\boldL) - \bar{\alpha}(\boldL)$ is in Figure~\ref{fig:birefmap}. The power spectra of the mean-field biases to both screening and cosmic birefringence are presented in Figures \ref{fig:tau_biases} and \ref{fig:rot_biases}. Because screening acts as a weak mask to the CMB photons, its quadratic estimator is similar to that for a general mask. Therefore, the screening estimator is especially sensitive to the real space mask, and thus has a noticeably large mean-field bias at large reconstructed scales, $L<100$. We therefore choose $L_\mathrm{min}^{\tau} = 100$ for this analysis because even though we thoroughly estimate the mean-field, its large amplitude means any uncertainty would have a strong impact on the residual band-powers below this scale.

\subsection{Power spectrum biases}\label{sec:psbiases}
Given a normalized, reconstructed map $\hat{\Gamma}$, and its estimated mean-field $\bar{\Gamma}$, we compute its power spectrum:
\begin{equation}
    C_{L\mathrm{, raw}}^{\hat{\Gamma} \hat{\Gamma}} = C_L^{\hat{\Gamma} \hat{\Gamma},\times}\left[\left(\hat{\Gamma}(\boldL) - \bar{\Gamma}(\boldL)\right), \left(\hat{\Gamma}(\boldL) - \bar{\Gamma}(\boldL)\right) \right]
    \label{eq: mfsub_ps}
\end{equation}
The raw reconstructed power spectrum contains several bias contributions beyond the target signal:
\begin{equation}
    \begin{split}
    C_L^{\hat{\Gamma} \hat{\Gamma}} = &\frac{1}{\mathrm{TF}} \cdot C_{L\mathrm{, raw}}^{\hat{\Gamma} \hat{\Gamma}}\\ & - \frac{1}{\mathrm{TF}} \cdot \left[ N_L^{0,\Gamma} + N_L^{1,\Gamma\phi} + N_L^{\mathrm{leak,\ }\Gamma\phi} + \mathrm{MC Bias}^{\Gamma}\right]
    \end{split}
    \label{eq:sum_recon}
\end{equation}
where $N_L^{0,\Gamma}$ is the disconnected Gaussian bias (Section~\ref{subsec:n0bias}), $N_L^{\mathrm{leak,\ }\Gamma\phi}$ is the primary lensing bias to $\Gamma$ (discussed in Section~\ref{subsec:n1andnleak}), $N_L^{1,\Gamma\phi}$ is the secondary lensing bias to $\Gamma$ (also in Section ~\ref{subsec:n1andnleak}), and $\mathrm{MC Bias}^{\Gamma}$ is any leftover bias from the simulation-based estimates (Section~\ref{subsec:adbias}). Because these biases are well-studied in CMB lensing analyses, we can similarly quantify and subtract the equivalent biases for the reconstructed screening and birefringence band-powers.


\subsubsection{$N_L^{0, \Gamma}$ Bias Subtraction}\label{subsec:n0bias}
The $N_L^{0, \Gamma}$ or Gaussian bias arises from random-chance correlations between CMB modes that mimic the mode-coupling signature of the target signal, even in the absence of screening or birefringence. It is the dominant bias at most reconstructed scales (Figures~\ref{fig:tau_biases} and~\ref{fig:rot_biases}). We estimate and subtract this bias using the realization-dependent $N^0$ (RDN0) algorithm \citep{dvorkinsmith2009,namikawa2013}, adapted for the cross-split estimator as described in Q24 (Appendix~E.1). This algorithm combines data (D) and simulation (S) maps in specific pairings in the split quadratic estimator (Equation \ref{eq:splitmaprecon_example}) to isolate the disconnected contractions of the four-point function. It has the useful property of being insensitive to small mismatches between the data and simulation power spectra. This result is obtained as:
\begin{multline}
    N_L^{0,\Gamma} = 2\left<\bar{C}_L^{\times}
    [\hat{\Gamma}^{\mathrm{D,\ S}}(\boldL),   \hat{\Gamma}^{\mathrm{D,\ S}}(\boldL)] \right.
    \\
    \left.
     -\  \bar{C}_L^{\times}
    [\hat{\Gamma}^{\mathrm{S,\ S'}}(\boldL),   \hat{\Gamma}^{\mathrm{S,\ S'}}(\boldL)]\ \right>_\mathrm{S,\ S'}
     \label{eq:n0bias}
\end{multline}
 We use 480 simulations from the Q24 suite for this calculation. The resulting $N_L^{0,\ \Gamma}$ biases for 1 test simulation (subbed in for the D superscript) are shown in Figures \ref{fig:tau_biases} and \ref{fig:rot_biases} for screening and birefringence, respectively. For our data results, we use the real data where the D superscripts appear in Equation~\ref{eq:n0bias} to obtain this bias. 

\subsubsection{Lensing-induced biases} \label{subsec:n1andnleak}
The remaining dominant biases arise from primary and secondary contractions of the CMB lensing field that contaminate the $\Gamma$ fields in the standard quadratic estimator.

\paragraph{Primary lensing leakage, $N_L^{\mathrm{leak,\ }\Gamma \phi}$} CMB lensing leaks into the anisotropic screening estimator because the effects couple off-diagonal CMB modes in a similar way. As can be seen from Table~\ref{tab:filters}, the lensing and screening $TT$ mode-couplings have the same structure, but the lensing version weighs each term by $\mathbf{L}\cdot\boldsymbol{\ell}_i$. Because CMB lensing is such a large effect compared to screening, we can think of the screening estimator as an inefficient lensing estimator. Therefore, lensing will certainly contaminate any estimate of screening. 

To handle this, one may use a lensing bias-hardened $\tau$ estimator but at the cost of increased error \citep[e.g.,][]{namikawa2013, namikawa2021}. However, we can reconstruct screening with no lensing bias-hardening, and instead construct an analytical curve for what we expect the lensing bias to be at the $C_L^{\hat{\tau}\hat{\tau}}$ level. Based on the derivation for a general bias-hardened quadratic estimator from \citet{namikawa2013}, the response function that describes how the $\tau$ estimator responds to the $\phi$ mode-coupling is as follows:
\begin{equation}
    \hat{\tau}(\boldL) = \tau(\boldL) + R_L^{\tau \phi}\,\phi(\boldL),
\end{equation}
where
\begin{equation}
R^{\tau \phi}_{L} = A_L^{\tau} \int \frac{d^2 \boldell_1}{(2\pi)^2} 
\frac{f^\tau(\boldell_1,\boldell_2)\, 
f^\phi(\boldell_1,\boldell_2)}
{2\,C_{\ell_1}^{TT,\mathrm{tot}}\,
C_{\ell_2}^{TT,\mathrm{tot}}}.
\label{eq:rl}
\end{equation}

Taking the power spectrum of $\hat{\tau}_L$ we find:
\begin{equation}
\begin{split}
    C_L^{\hat{\tau}\hat{\tau}} &= C_L^{\tau \tau,\ \mathrm{theory}} + (R_L^{\tau \phi})^2 \cdot C_L^{\phi \phi,\ \mathrm{theory}}\\& \equiv C_L^{\tau \tau,\ \mathrm{theory}} + N_L^{\mathrm{leak,}\tau \phi},
    \label{eq:nlleak}
\end{split}
\end{equation}
where in the second line we defined $N_L^{\mathrm{leak,}\tau \phi}$. Note that we neglected the theory cross-term between $\tau$ and $\phi$ here; using the theory curve from \citet{FengHolder2019}, we found the leakage from this term to be $\sim$ two orders of magnitude smaller than our screening errors. The $N_L^{\mathrm{leak,}\tau \phi}$ lensing bias for the ACT DR6 analysis is shown in Figure~\ref{fig:tau_biases} compared to the other biases discussed in this section. We compute this analytically using Equation~\ref{eq:nlleak}, but find good agreement with our simulations. We note that this lensing bias is only calculated for screening because the $EB$ mode-couplings for birefringence and lensing are orthogonal at map-level \citep{kamionkowski2009}. However, birefringence and lensing do induce higher-order biases on each other's reconstructed power spectra \citep{cai2023, cai2024}. Therefore, both screening and birefringence pick up a secondary lensing contamination as discussed in the next paragraph.

\paragraph{Secondary lensing contamination, $N_L^{1,\Gamma\phi}$} The biases from secondary contractions of anisotropic fields entering the reconstructed $\Gamma$ power spectrum, beyond the primary contraction that the estimator is designed to measure, are denoted generally as $N_L^{1, \Gamma\phi}$. For instance, Figure 1 in \citet{kesden2003} depicts quadrilaterals formed by CMB modes $T(\boldell)$, whose diagonals are the desired reconstructed lensing modes $\phi(\boldL)$ and the perpendicular lensing modes not of $\boldL$ that additionally contribute to the power measured at scale $\boldL$. The sum of these additional contractions yield this so-called ``connected'' bias. These connected biases notably contain biases from fields other than $\Gamma$ into the desired $\Gamma$ reconstruction. In this case, our main simulation suite does not contain the $\Gamma$ fields but \emph{does} contain lensing, which is known to percent-level, and can be safely estimated and removed. Thus, our $N_L^{1,\phi \Gamma}$ for $\Gamma \in [\tau, \alpha]$ is the estimate of the secondary connected \emph{lensing} bias to their reconstructed power spectra. The $N_L^{1,\phi\alpha}$ bias is discussed in detail in \citet{cai2024}, where they show that even though lensing and birefringence do not bias each other at the reconstructed map-level, they do contaminate each other at the power spectrum-level. 

We estimate this bias following Q24 (Appendix~E.2), using two categories of noiseless lensed CMB simulation pairs, 90 pairs each. The first category, denoted $(S_\phi, S'_\phi)$, consists of pairs that share a common lensing potential $\phi$ but have independent CMB realizations. The second category, denoted $(S, S')$, consists of pairs with fully independent CMB \emph{and} lensing realizations. The $N_L^{1,\Gamma\phi}$ bias is then
\begin{equation}
\begin{split}
    N_L^{1,\phi \Gamma} = \langle\, 
    & \bar{C}_L^{\times}
    [\hat{\Gamma}^{\rm{S}_\phi, \rm{S}'_\phi}(\boldL),\, 
    \hat{\Gamma}^{\rm{S}_\phi, \rm{S}'_\phi}(\boldL)] \\
    +\, & \bar{C}_L^{\times}
    [\hat{\Gamma}^{\rm{S}_\phi, \rm{S}'_\phi}(\boldL),\, 
    \hat{\Gamma}^{\rm{S}'_\phi, \rm{S}_\phi}(\boldL)] \\
    -\, & \bar{C}_L^{\times}
    [\hat{\Gamma}^{\rm{S}, \rm{S'}}(\boldL),\, \hat{\Gamma}^{\rm{S}, \rm{S'}}(\boldL)] \\
    -\, & \bar{C}_L^{\times}
    [\hat{\Gamma}^{\rm{S}, \rm{S'}}(\boldL),\, \hat{\Gamma}^{\rm{S'}, \rm{S}}(\boldL)]
    \,\rangle_{S,S',S_\phi,S'_\phi},
\end{split}
\end{equation}
where superscripts indicate which simulation set each input field is drawn from.

\subsubsection{Additive MC Bias}\label{subsec:adbias}
After subtracting the $N_L^{0,\Gamma}$, $N_L^{\mathrm{leak,}\tau\phi}$, and $N_L^{1,\phi \Gamma}$, a small residual bias remains from higher-order terms in the quadratic estimator and from imperfections in the bias estimates themselves (e.g. approximate treatment of masking). Following Appendix~E.3 of Q24, we absorb these residuals into an additive Monte Carlo correction. Because the Q24 simulations contain no screening or birefringence signal, the expectation value of the fully-debiased simulation band-powers is zero. Any nonzero remainder is the MC bias:
\begin{equation}
    \mathrm{MCBias}^\Gamma = 
    \left\langle C_L^{\hat{\Gamma}\hat{\Gamma},\times}
    \right\rangle_{\mathrm{sim}} 
    - \mathrm{MC}N_L^{0,\Gamma} 
    - N_L^{1,\Gamma} 
    - N_L^{\mathrm{leak,}\Gamma\phi}.
    \label{eq:mcbias}
\end{equation}
This is subtracted from the data band-powers \emph{before} the transfer function correction in Equation~\ref{eq:sum_recon}, so no $1/\mathrm{TF}$ pre-factor appears here. The $\mathrm{MC}N_L^{0,\Gamma}$ term is the simulation-only analogue of the realization-dependent $N_L^{0,\Gamma}$: both legs of both reconstructions are drawn from simulations, with no data involved. Following the notation of Q24 (Appendix~E.1),
\begin{equation}
\begin{split}
    \mathrm{MC}N_L^{0,\Gamma} = \Big\langle\,
    &C_L^{\times}\!\left[
    \hat{\Gamma}^{\rm{S},\rm{S'}}(\boldL),\, \hat{\Gamma}^{\rm{S},\rm{S'}}(\boldL)\right] \\
    &+ C_L^{\times}\!\left[
    \hat{\Gamma}^{\rm{S},\rm{S'}}(\boldL),\, \hat{\Gamma}^{\rm{S'},\rm{S}}(\boldL)\right]
    \Big\rangle_{\rm{S},\rm{S}'},
\end{split}
\end{equation}
where $S$ and $S'$ denote independent simulation pairs. We use 480 pairs for this estimate, and subtract this quantity from the average of the simulation reconstructions to obtain the MC Bias as shown in Equation~\ref{eq:mcbias}. 

\subsection{Band-power Covariances}\label{sec:covariance}
We estimate the covariance matrices $\mathcal{C}^\Gamma$ for screening and birefringence from the scatter of 792 $\Gamma$ reconstructions of the Q24 simulation suite; the corresponding correlation matrices are shown in Figure~\ref{fig:covmats} in the Appendix. We expect this to be a sufficient number of simulations given our number of bins \citep[e.g.,][]{hartlap2007,sellentin2016}. For screening, adjacent bins at high $L$ are strongly correlated, but because the constraining power on $A_\tau$ comes from lower $L$ (see Figure~\ref{fig:filters}), we can choose a binning that keeps the most informative bins approximately independent without degrading the final constraint. In the case of birefringence, the correlation matrix is generally well-behaved. After confirming that their conditions are acceptable, we invert these matrices to obtain the band-power errors on each measurement.
\subsection{Pipeline Verification}\label{sec:implement}
\begin{figure}[htp!]
    \centering
    \includegraphics[width=0.9\linewidth]{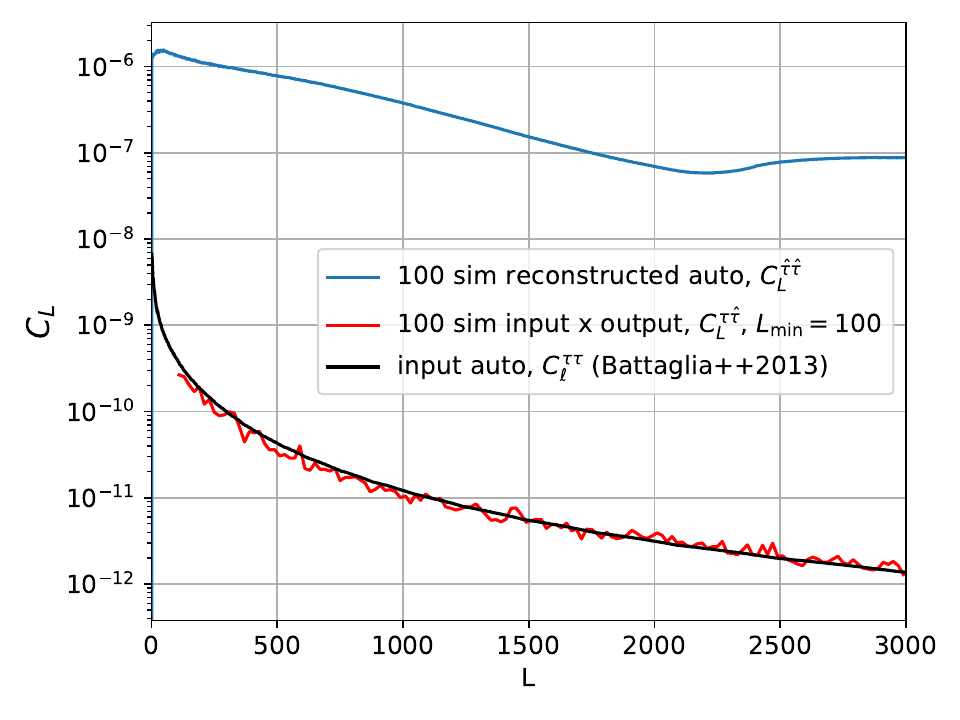}
    \hspace{-0.4cm}
    \includegraphics[width=\linewidth]{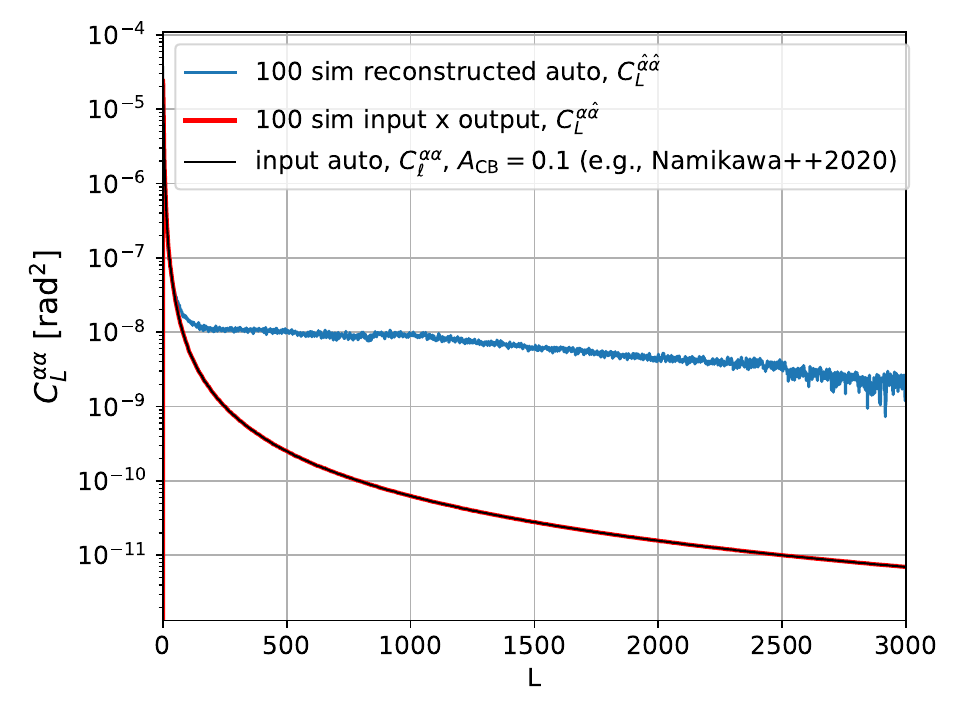}
    \caption{Results of a pipeline verification test on both screening-modulated and birefringence-rotated CMB simulations. These consist of input (black) and reconstructed (blue) auto-power spectra with the input/output cross-power spectrum on top (red). These results validate our cross-split reconstruction methods because the input-cross-output spectra agree very closely in shape and amplitude to the respective input spectra. \emph{Top:} Reconstructed, input, and input-cross-output spectra for screening. The reconstructed and cross spectra are the average of 100 spectra made with simulations that each have unique $\tau(\hat{\mathbf{n}})$ realizations. Note that the in-cross-out, binned with $\Delta L = 20$ here, traces the input well but has high scatter due to the large amount of reconstruction noise (as shown in the difference between the blue and black curves). \emph{Bottom:} The same as the top panel but for anisotropic cosmic birefringence. This input-cross-output result has less scatter due to the reconstructed autospectrum being much closer to the input, which means the reconstruction noise is much lower, relatively speaking, than for $\tau$.} 
    \label{fig:pipever}
\end{figure}
To validate these newly-implemented estimators, we use the 100 signal-injected simulations described in Section~\ref{sec:sims}. We pass each simulation through the mean-field estimation and reconstruction pipelines, and cross-correlate the output map with the known input signal to assess signal recovery. Figure~\ref{fig:pipever} shows the input auto-spectrum, the reconstructed auto-spectrum, and the input-output cross-spectrum for both $\tau$ and $\alpha$. In both cases, the cross-spectrum traces the input closely, confirming that the pipeline recovers the target signal. The reconstructed auto-spectra are much larger than the input because reconstruction noise dominates over these weak signals, even in the absence of instrumental noise.

To validate the entire power spectrum pipeline, we treat one Q24 simulation as the data and pass it through the pipeline. We calculate and remove each bias from Equation~\ref{eq:sum_recon}, paying particular note to the realization dependence of the $N_L^{0, \Gamma}$ bias. The resulting simulation-based band-powers are shown in Figures~\ref{fig:tau_biases} and \ref{fig:rot_biases}. We find that these single-simulation band-powers are consistent with zero; their probabilities-to-exceed (PTEs), otherwise known as $p$-values, are 0.73 for screening (with 21 degrees of freedom) and 0.89 for birefringence (with 19 degrees of freedom).

\section{Results}\label{sec:results}

We apply the methods laid out in Section~\ref{sec:methods} to the DR6 data (described in Section~\ref{sec:data}) according to the screening and birefringence formalism, including removing all relevant biases as discussed in Sections \ref{sec:mapbiases} and \ref{sec:psbiases}, and present our results below.

\subsection{Anisotropic Screening}\label{sec:tau_results}

\begin{figure*}[htp!]
    \centering
    \includegraphics[width=0.85\linewidth]{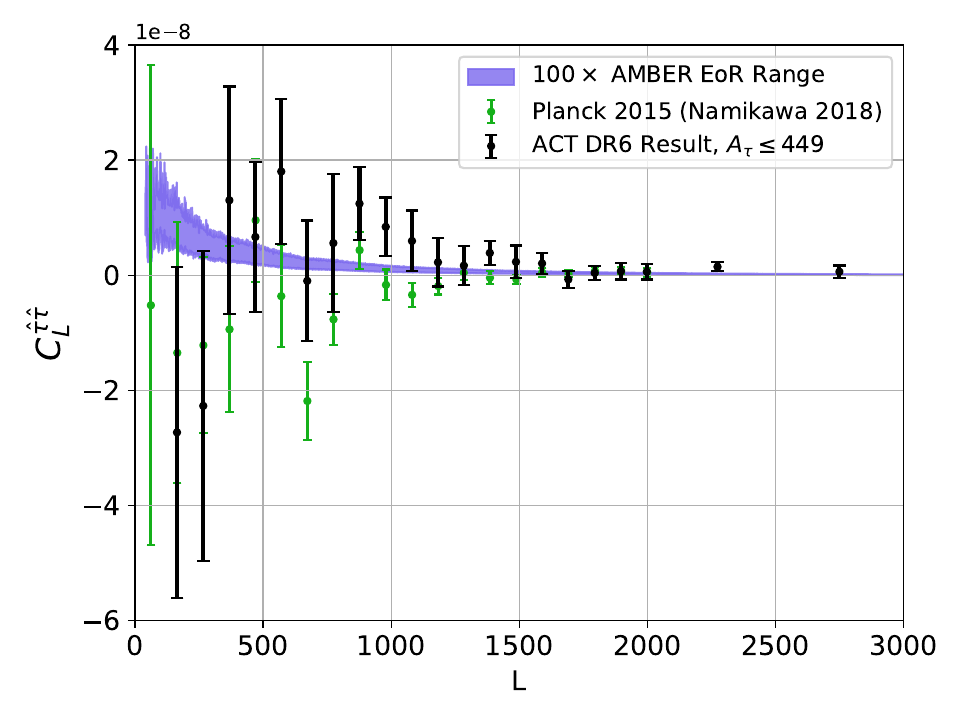}
    \caption{Presentation of the ACT DR6 baseline screening band-powers for $100 < L^\tau < 3000$ (black) in comparison to the \textsl{Planck} 2015 \citet{toshiyatau2018} data (light green). The purple band represents a range of plausible \texttt{AMBER} theory curves multiplied by 100. Not shown here are the \textsl{WMAP} \citep{gluscevic2012} and BICEP/Keck \citep{Ade2023} screening band-powers because they exceed the vertical range of this plot.}
    \label{fig:tau_bps}
\end{figure*}

We present constraints on the screening power spectrum, reconstructed using the $TT$ quadratic estimator, in Figure~\ref{fig:tau_bps}. For this result, we choose $L_\mathrm{min}^\tau = 100$ due to the large mean-field bias described in Section~\ref{subsec:mfbias}. We choose $L^\tau_\mathrm{max} = \ell_\mathrm{CMB,max} = 3000$ for these band-powers because 1) our tests show that foregrounds do not have any significant effect at high reconstructed $L^\tau$ (see Figure~\ref{fig:coadd_fgsum}), and 2)  $N_L^{0,\tau}$ increases sharply for $L^\tau > \ell_\mathrm{CMB,max}$. Our result is consistent with a null signal, yielding a $\chi^2$ whose PTE is 0.68 (see Table \ref{tab:consistency_tests}) for 21 degrees of freedom (dof).

\begin{figure}[htp!]
    \centering
    \includegraphics[width=\linewidth]{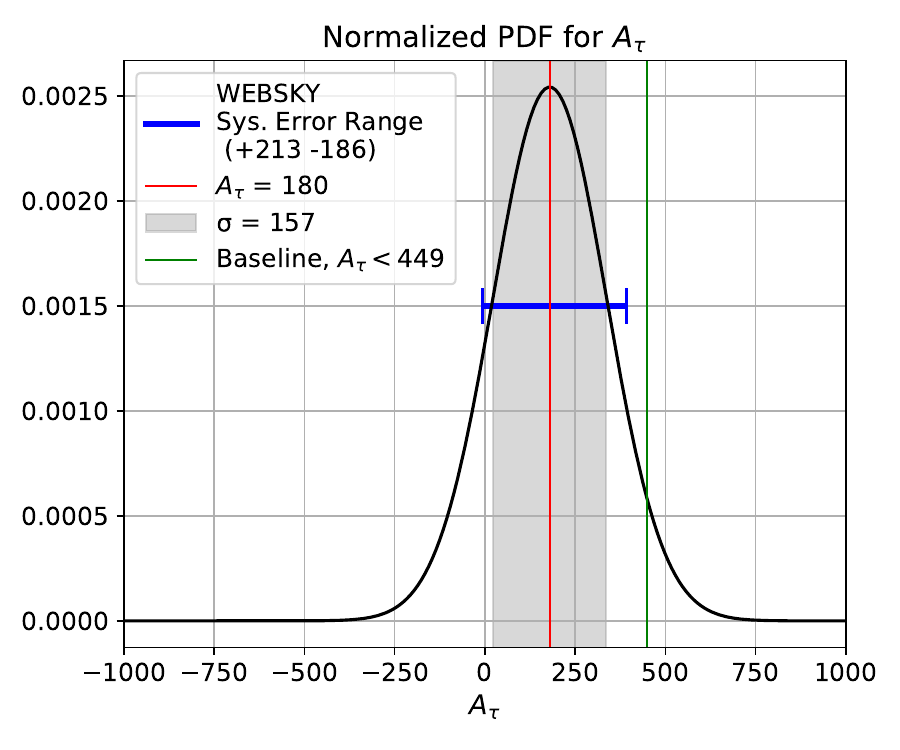}
    \caption{Normalized posterior from fitting an amplitude of the fiducial \texttt{AMBER} $C_\ell^{\tau \tau}$ to the ACT DR6 band-powers. We find $A_\tau = $\atau, or an upper limit of $A_\tau \lesssim \ $\atauUL\ at 95\% confidence (excluding zero from the integral). This result is orders of magnitude away from detecting a signal with the fiducial amplitude of $A_\tau = 1$, but is the first upper limit on an \texttt{AMBER}-based reionization model. We also show a blue horizontal error representing the estimated systematic error bar based on the \texttt{WEBSKY} model. This demonstrates that foreground systematics are likely significant in this analysis, which we discuss in detail in Section \ref{subsec:fgs}.}
    \label{fig:taupost}
\end{figure}

We also show the recent band-powers and errors on this power spectrum from \textsl{Planck} 2015 data \citep{toshiyatau2018}. Other measurements not shown here are from \textsl{WMAP} \citep{gluscevic2012} and BICEP/Keck \citep{Ade2023}. All band-power-level screening constraints thus far have found no hint of a signal, indicating that this method would benefit from more sensitive data. For a theory reference, we have included a representative range of reionization models obtained from \texttt{AMBER} simulations \citep{trac2022, Chen2023} \emph{multiplied by $\mathit{100}$}. To help estimate our overall uncertainty, we use a theoretical scaling of a template for $C_L^{\tau\tau}$ based on the fiducial $\tau$ power spectrum from the \texttt{AMBER} simulations:
\begin{equation}
C_L^{\tau \tau} =
A_\tau \, C_L^{\tau \tau,\mathrm{fid}}.
\end{equation}
This assumed patchy screening power spectrum depends on various reionization parameters whose names and fiducial values are the following: [$z_{\text{mid}} = 8$, $\Delta z = 4$, $A_z = 3$, $M_{\text{min}} = 10^8 M_{\odot}$, $l_{\text{mfp}} = 3$ Mpc/h]. Please see \citet{trac2022} and \citet{Chen2023} for more information on these simulations and their parameters.

We now describe the process of fitting our results as a function of  $A_\tau$. The $\chi^2$ distribution for the data $C_L^{\hat{\tau}\hat{\tau}}$, covariance matrix $\mathcal{C^\tau}$, and model $C_L^{\tau \tau,\mathrm{fid}}$ given here is:
\begin{equation}
    \begin{split}
    & \chi^2 (A_\tau) \\ & = \sum_{LL'}\left(C_L^{\hat{\tau}\hat{\tau}} - A_\tau C_L^{\tau \tau,\mathrm{fid}}\right)(\mathcal{C}^\tau)^{-1}_{LL'} 
    \left(C_{L'}^{\hat{\tau}\hat{\tau}} - A_\tau C_{L'}^{\tau \tau,\mathrm{fid}}\right).
    \end{split}
\end{equation}
We implement a likelihood of the form $  \mathcal{L} \propto \exp[-\chi^2/2].$

Given a flat prior on $A_\tau$, we solve Bayes' Theorem to find the one-dimensional probability distribution for $A_\tau$ shown in Figure~\ref{fig:taupost}. We find $A_\tau = $\atau, or an upper limit of $A_\tau <$ \atauUL\ at 95\% confidence after requiring $A_\tau \geq 0$. This is a result consistent with null, alongside other recent measurements \citep[e.g.,][]{toshiyatau2018, Ade2023}. This is very unconstrained due to the $\sim 1000\times$ factor between the DR6 noise level and the fiducial $\tau$ signal. This type of upper limit provides a starting point for future surveys to place constraints on simulation-generated models.

\subsection{Anisotropic Cosmic Birefringence} \label{subsec:biref_results}

\begin{figure*}[!h]
    \centering
    \includegraphics[width=0.85\linewidth]{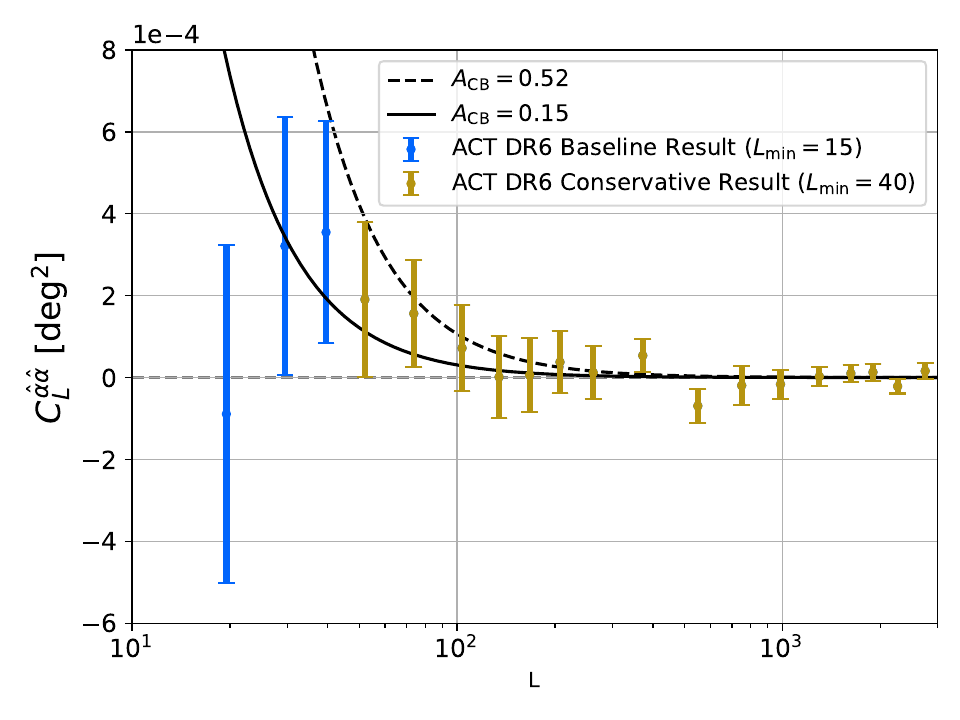}
    \caption{ACT DR6 anisotropic cosmic birefringence band-powers and the associated errors. These band-powers have been bias-subtracted as discussed in Section \ref{sec:methods}. The two colors of band-powers represent the two results we report: The blue set, with $L^\alpha_\mathrm{min} = 15$, are from our baseline analysis which represent $A_\mathrm{CB} <$ \aCBUL\ (the solid theory curve) at 95\% confidence. The gold set of band-powers, which are a subset of the blue but have $L^\alpha_\mathrm{min} = 40$, represent $A_\mathrm{CB} < 0.52$ (the dashed theory curve) at 95\% confidence.}
    \label{fig:biref_bps}
\end{figure*}

We show the bias-subtracted anisotropic cosmic birefringence $EB$ band-powers and error bars from ACT DR6 in Figure~\ref{fig:biref_bps}, with their correlation matrix (calculated the same way as for screening) shown in the right panel of Figure~\ref{fig:covmats}. We use two different values of $L_\mathrm{min}^\alpha$, 15 and 40, depending on the choice for handling the global angle mean-field bias (described in Section~\ref{subsec:rot_mfbias}
). In both results, we choose $L_\mathrm{max}^\alpha = 3000$ because $N_L^{0,\alpha}$ increases sharply for $L_\mathrm{max}^\alpha > \ell_\mathrm{CMB,max}$.

\begin{figure}[!h]
    \centering
    \includegraphics[width=\linewidth]{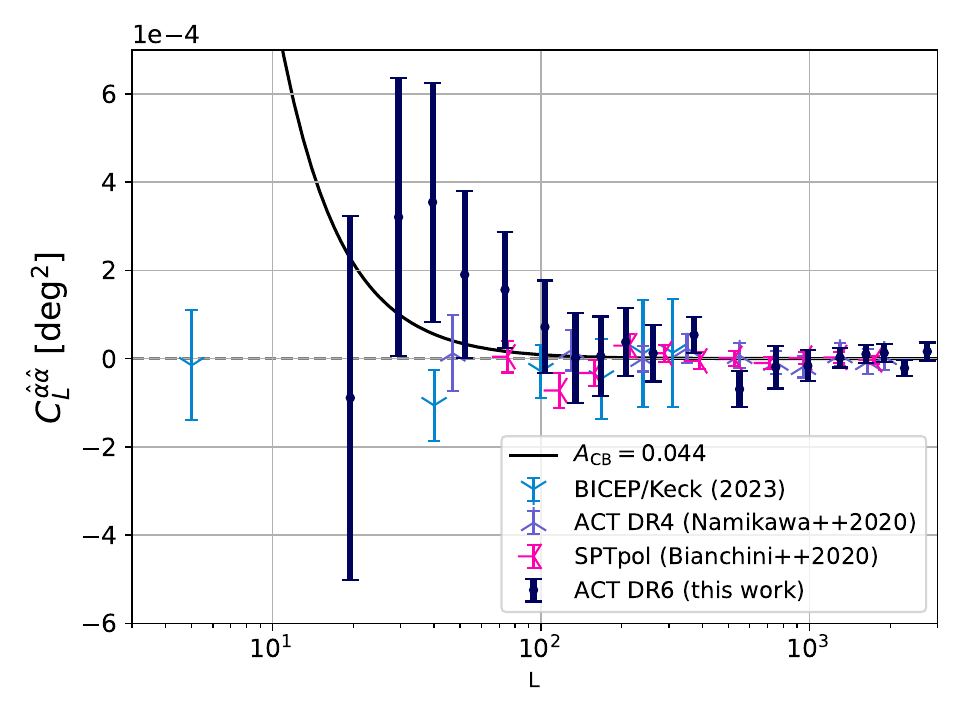}
    \caption{Comparison of our ACT DR6 cosmic birefringence band-powers (in dark blue) to other recent ground-based CMB survey results; BICEP/Keck (light blue), ACT DR4 (dark periwinkle) and SPTpol (hot pink). We do not show the recent \textsl{Planck} band-powers here because they are more finely binned and exist on larger reconstructed scales than these ground-based results. See \citet{Gruppuso_2020}, \citet{Bortolami_2022}, and \citet{zagatti2024} for the \textsl{Planck} birefringence band-powers. Of those shown here, BICEP/Keck place the tightest upper limit on $A_\mathrm{CB}$, with ACT DR4 and SPT closely following. We note that ACT DR6 has slightly less constraining power on this measurement than the ACT DR4 analysis. This is due to the use of many more large-scale CMB modes in the ACT DR4 data, down to $\ell_\mathrm{min} = 200$, as well as the reduced polarization noise in ACT DR4 compared to ACT DR6.}
    \label{fig:biref_bps_comp}
\end{figure}

\begin{figure}[!h]
    \centering
    \includegraphics[width=\linewidth]{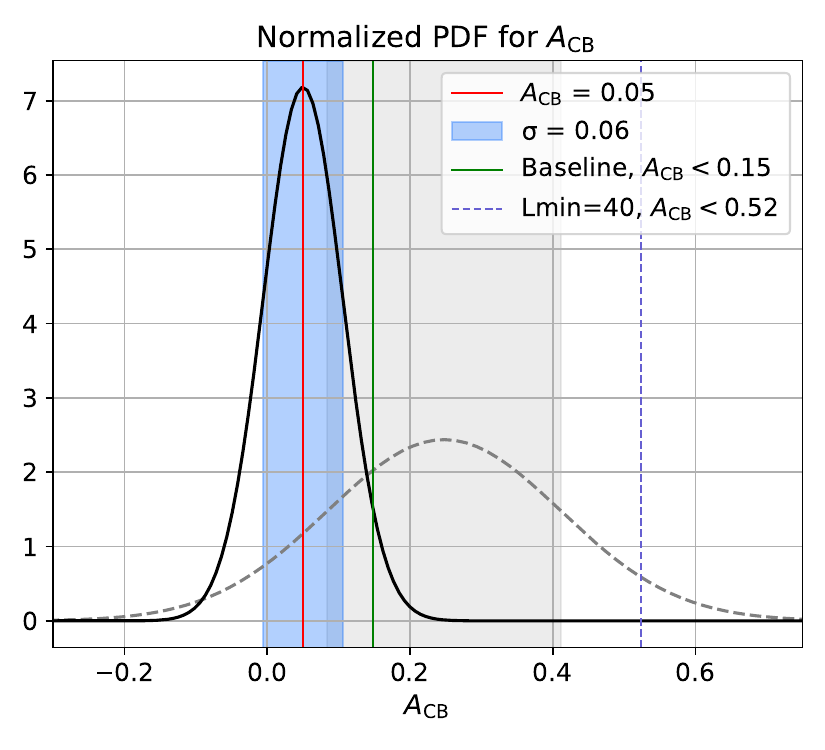}
    \caption{The normalized posterior for our baseline measurement of $A_\mathrm{CB} =$\ \aCB\  ($1\sigma$) in black. We find a 95\% confident upper limit (truncated from below at 0 to disallow negative values) of $A_\mathrm{CB}<$\ \aCBUL\ (green vertical line). This is consistent with null and with the other current upper limits on the amplitude of anisotropic birefringence. We also present the more conservative result discussed in Section \ref{subsec:rot_mfbias} in gray, representing an upper limit of $A_\mathrm{CB} < 0.52$ (dashed periwinkle line).}
    \label{fig:birefpost}
\end{figure}

We again calculate a Gaussian likelihood with a flat prior on $A_\mathrm{CB}$, in the same exact manner as for screening but substituting the amplitude $A_\mathrm{CB}$ and the fiducial anisotropic cosmic birefringence power spectrum $C_L^{\alpha \alpha}$ (defined in Equation \ref{eq:acbdef}). Using $L_\mathrm{min}^\alpha = 15$, we find $A_\mathrm{CB} =$ \aCB, or an upper limit of $A_\mathrm{CB} <$ \aCBUL\ at 95\% confidence after requiring $A_\mathrm{CB} \geq 0$. This result is consistent with null and has a PTE of 0.75 (see Table \ref{tab:consistency_tests}) with 19 dof. This result is also consistent with all previous anisotropic birefringence studies which find $A_\mathrm{CB}$ consistent with zero. From Equations~\ref{eq:claadef} and \ref{eq:acbdef}, this baseline $A_\mathrm{CB}$ result represents an upper limit on the Chern-Simons term $g_{a \gamma} <\ $\gUL.

This result is dependent on our analysis of the global polarization angle, therefore we also present a more conservative upper limit of $A_\mathrm{CB} < 0.52$; see Section \ref{subsec:rot_mfbias} for details. In either case, the DR6 result is less constraining than other current ground-based measurements \citep[e.g.,][]{namikawa2020,Bianchini_2020,Ade2023}, shown alongside our measurement in Figure~\ref{fig:biref_bps_comp}. This is primarily because the DR6 polarization noise and the $\ell_\mathrm{min} = 600$ restriction reduce access to the large CMB scales where the birefringence estimator is most sensitive (Figure~\ref{fig:filters}). Furthermore, our use of the cross-split estimator, while more robust to noise biases, increases the variance of our final measurement compared to a standard ``coadd'' quadratic estimator \citep{madhavacheril2020}. This penalty is more severe here than in the DR6 lensing analysis, which is dominated by the temperature estimator: discarding the auto-noise contributions costs more when the legs are noise-dominated, as the $B$ modes are at all scales used here.

\section{Systematics Tests} \label{sec:systematics}
As shown in Sections~\ref{sec:mapbiases} and \ref{sec:psbiases}, we quantify and subtract many well-understood biases to these reconstructions. However, any unknowns that are in the data have the potential to cause systematic biases that we did not quantify. We consider screening foregrounds, a separate birefringence mean-field bias, and a suite of robustness tests that validate analysis choices. 

\subsection{Screening Foregrounds}\label{subsec:fgs}
Because screening is reconstructed via a quadratic estimator of the CMB temperature data in this work, this measurement is susceptible to higher-order functions of the CMB temperature foregrounds. These foregrounds include cosmic infrared background (CIB) emission, radio galaxy emission, and thermal Sunyaev-Zel'dovich (tSZ) contamination. It is well-understood that foregrounds can have significant effects on CMB lensing and other four-point CMB analyses due to their non-Gaussian nature \citep[e.g.,][]{vanengelen2014,osborne2014,maccrann2024,maccrann2024a}, so it is important to investigate the effect of foregrounds here as well. To estimate the possible effect of residual non-Gaussian foregrounds in the data, we use the \texttt{WEBSKY} \citep{stein2020} residual foreground simulations that were processed and studied in \citet{maccrann2024} for the lensing results of Q24. To get self-consistent residual foreground maps, the 90 and 150 GHz \texttt{WEBSKY} CIB, tSZ, kSZ, and radio maps were masked for point-sources and galaxy clusters, then coadded using the frequency weights from the DR6 lensing analysis. This process mimics the processing of the DR6 data used in this work, and thus is the representation of the foreground contamination that would be in the data based on the \texttt{WEBSKY} models. We then pass the resulting 90 GHz, 150 GHz, and frequency-coadded simulations (and in the case of the bispectra, the \texttt{WEBSKY} lensed CMB realization) through the coadd quadratic estimator for screening, estimate and subtract their mean-field and $N_L^0$ biases, and obtain the reconstructed residual foreground bias power spectrum. 

\begin{figure}
    \includegraphics[width=\linewidth]{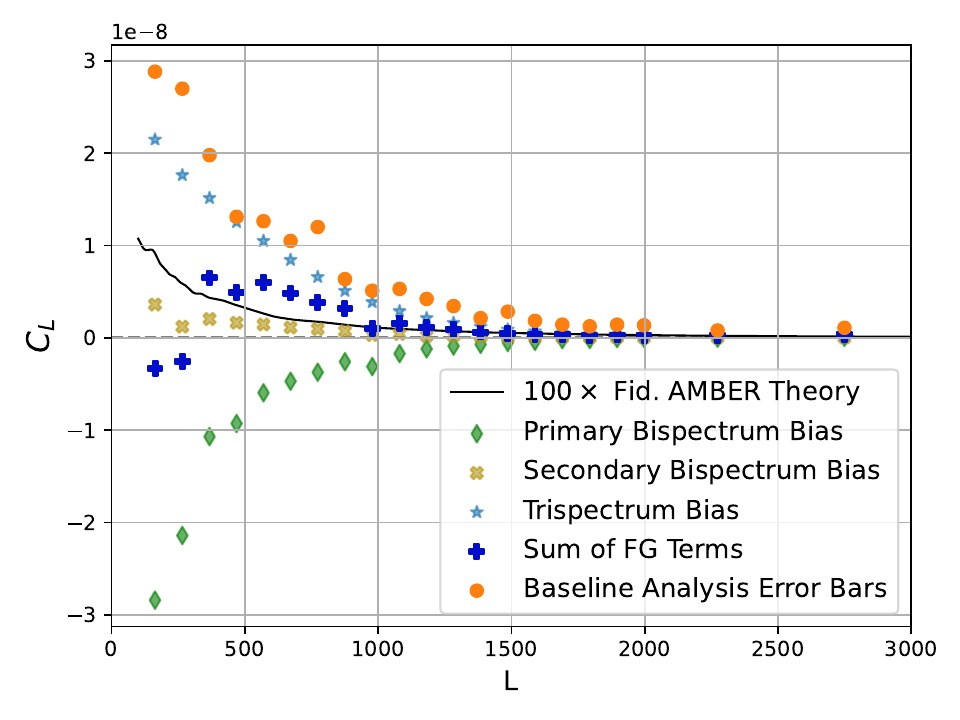}
    \caption{A comparison of the total residual foreground contamination according to \texttt{WEBSKY} (purple plus signs), to the baseline statistical screening errors in this work (orange circles). Also shown are the individual \texttt{WEBSKY}-based foreground terms in green diamonds, gold x's, and light blue stars (see Equation~\ref{eq:fgterms} for each term). These results demonstrate that the potential systematic foreground bias is significant compared to the statistical errors in this analysis. The black line represents $100\times$ the fiducial \texttt{AMBER} signal, demonstrating that the possible strength of this foreground contamination is dominant over the desired signal.}
    \label{fig:coadd_fgsum}
\end{figure}

\begin{figure*}
    \centering
    \includegraphics[width=0.495\textwidth]{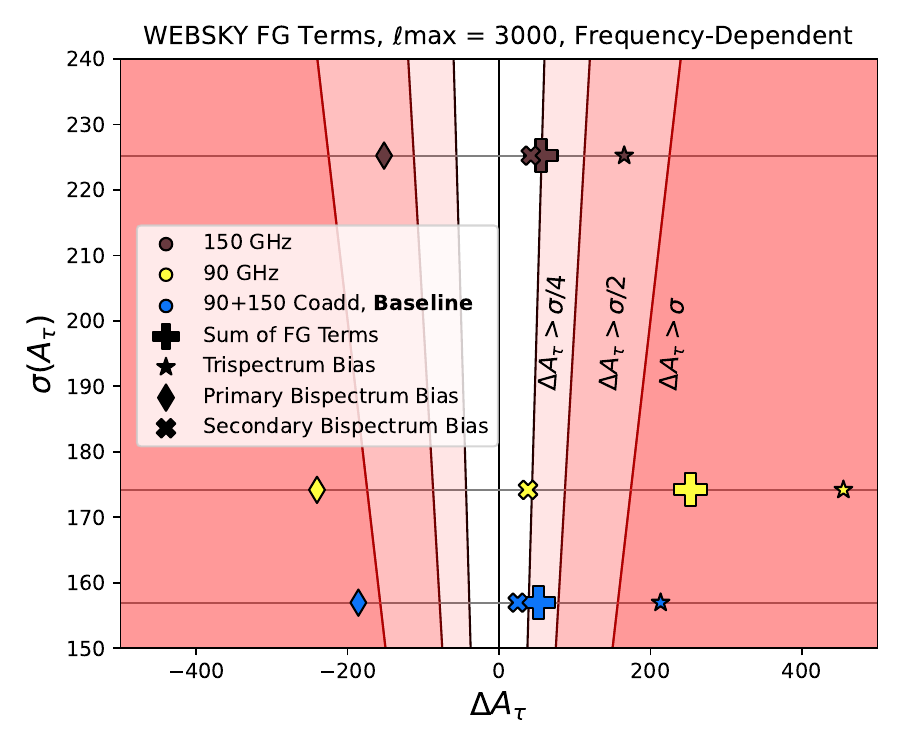}
    \includegraphics[width=0.495\textwidth]{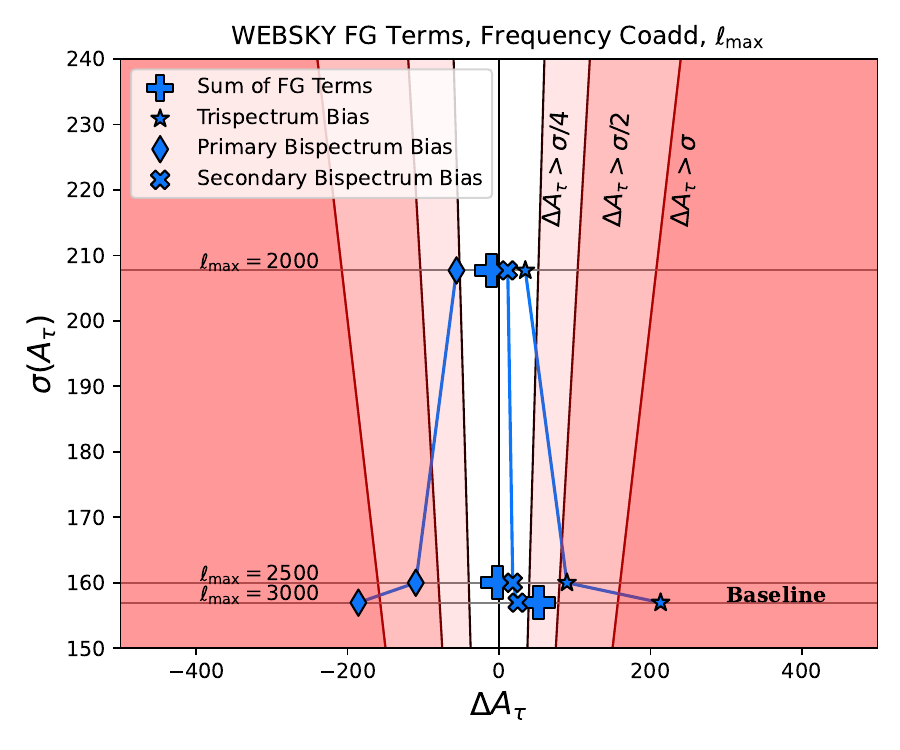}
    \caption{\emph{Left:} The estimated systematic bias to $A_\tau$, $\Delta A_\tau$, relative to the statistical error $\sigma(A_\tau)$ from each residual foreground term (various symbols) and their sum (heavy plus signs) with $\ell_\mathrm{max}$ fixed to 3000, presented for 90 GHz (yellow), 150 GHz (brown), and their coadd (blue). We find that the potential bias to the amplitude of screening is by far the largest coming from contaminants in the 90 GHz channel, likely due to residual tSZ signal. Specifically, we calculate the sum of the foreground terms $\Delta A_\tau = 0.33\sigma$ for the frequency coadd, $\Delta A_\tau =0.24\sigma$ for 150 GHz, and $\Delta A_\tau =1.4\sigma$ for 90 GHz with respect to the error bars from each frequency analysis. \emph{Right:} The systematic bias $\Delta A_\tau$ compared to the statistical uncertainty $\sigma (A_\tau)$ for the frequency coadd case with varying $\ell_\mathrm{max}$ of 3000, 2500, and 2000. This plot demonstrates that while for the baseline $\ell_\mathrm{max} = 3000$ case the systematic error is significant compared to the statistical error, reducing $\ell_\mathrm{max}$ both decreases the systematic error range, $\Delta A_\tau$, and increases the statistical error, $\sigma(A_\tau)$.}
    \label{fig:websky}
\end{figure*}

Specifically, we use the foreground residual simulations to reconstruct each component of the foreground contamination to the quadratic estimator, as in Equation 6 of \citet{maccrann2024a} and presented in, e.g., \citet{schaanferraro2019}. The change in the measured $\tau$ power spectrum from residual foregrounds can be written as:
\begin{multline}\label{eq:fgterms}
    \Delta C_L^{\hat{\tau}\hat{\tau}} = \\
    2C_L^\times [\hat{\tau}^{T_\mathrm{FG}, T_\mathrm{FG}}(\boldL),\ \hat{\tau}^{T_\mathrm{CMB},T_\mathrm{CMB}}(\boldL)] \\
    + 4C_L^\times [\hat{\tau}^{T_\mathrm{FG}, T_\mathrm{CMB}}(\boldL),\ \hat{\tau}^{T_\mathrm{FG},T_\mathrm{CMB}}(\boldL)] \\
    + C_L^\times [\hat{\tau}^{T_\mathrm{FG}, T_\mathrm{FG}}(\boldL),\ \hat{\tau}^{T_\mathrm{FG},T_\mathrm{FG}}(\boldL)]\\
\end{multline}
where $T_\mathrm{CMB}$ represents the lensed, but unscreened CMB temperature map, $C_L^\times$ represents taking the cross-power spectrum of the given maps, and the $\hat{\tau}(\boldL)$ represents the coadd screening estimator. The three terms on the right-hand side represent the primary and secondary bispectrum terms and the foreground trispectrum, listed in that order. Because the published \texttt{WEBSKY} suite does not contain maps of screening, there is no screening in the CMB maps to obtain the bispectrum terms. We therefore use the same simulations and methods for this calculation as \citet{maccrann2024a} but with the screening estimator instead of CMB lensing. The sum of these terms for the baseline case (frequency-coadd, $\ell_\mathrm{max} = 3000$) is compared to the data band-powers in Figure~\ref{fig:coadd_fgsum}.
To quantify the impact of these residual foregrounds, we fit the foreground bias spectrum from each term in Equation~\ref{eq:fgterms} to the fiducial screening template, obtaining a shift $\Delta A_\tau^\mathrm{FG}$ defined by $\Delta C_L^{\hat{\tau}\hat{\tau}} = \Delta A_\tau^\mathrm{FG}\, C_L^{\tau\tau,\mathrm{fid}}$. The uncertainty on $A_\tau$ versus the shift in $A_\tau$ from each residual \texttt{WEBSKY} foreground term in Equation \ref{eq:fgterms} for the various frequency and $\ell_\mathrm{max}$ choices tested here are shown in Figure~\ref{fig:websky}. Among the several analysis choices here, the sum of these foreground terms for the frequency coadd case with $\ell_\mathrm{max} = 3000$ (our baseline analysis choices) yields an estimated bias $\Delta A_\tau = 0.33\sigma$. This figure demonstrates a few other things: 

\begin{enumerate}
    \item The sum of all components for a given frequency choice and $\ell_\mathrm{max} = 3000$, according to \texttt{WEBSKY}, is expected to yield a positive bias to $A_\tau$, ranging from $0.24\sigma - 1.4\sigma$ for the three frequency cases. The largest bias comes from the 90 GHz channel.

    \item However, the secondary bispectrum term in these cases yields a negative bias which cancels some of the bias from the other two terms in the sum mentioned above. 

    \item Finally, for the coadded frequency case, reducing $\ell_\mathrm{max}$ improves the bias to $A_\tau$ compared to the $\ell_\mathrm{max} = 3000$ case at the expense of reduced sensitivity. 
\end{enumerate}
In general, these points indicate that foregrounds have the potential to contribute a non-neglilgible bias to our screening measurement. This is likely why the 90~GHz contamination is large enough to be visible in our single-frequency consistency test (Section~\ref{sec:const_tests}, Figure~\ref{fig:tau90150}).

\subsection{Effects of a global polarization angle on anisotropic cosmic birefringence}\label{subsec:rot_mfbias}
As explored in \citet{namikawa2020}, if there is a residual, uniform polarization angle on the whole sky or map, whether due to the instrument miscalibration or isotropic cosmic birefringence, or otherwise, it would contribute to the monopole of the anisotropic cosmic birefringence power spectrum via a global, parity-violating $C_\ell^{EB}$ spectrum. Due to the mask we apply, this monopole would leak into the smaller scales, similarly to a mean-field bias. 
\begin{figure}
    \centering
    \includegraphics[width=1\linewidth]{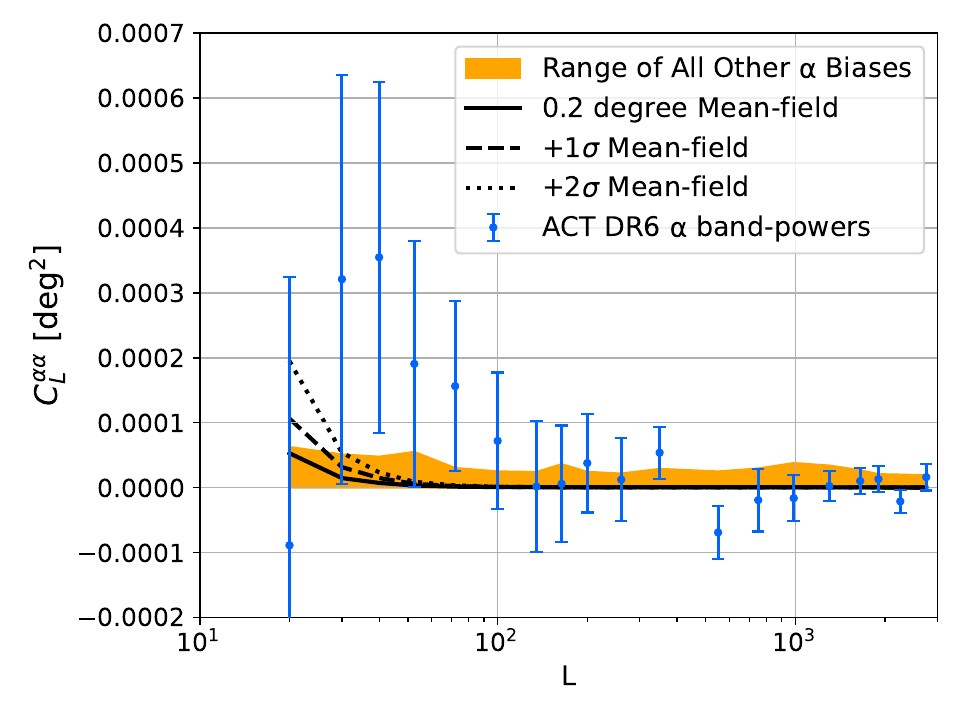}
    \caption{The mean-field bias expected to be present in the ACT DR6 birefringence reconstruction due to a global rotation angle in the CMB data, compared to the range of the other birefringence biases in this analysis and our baseline birefringence band-powers (black curves). Using the DR6 value found in \citet{louis2025} and its uncertainty, we show the three resulting binned mean-field  biases from the measurement (solid black) and its +1 (dashed black) and +2$\sigma$ (dotted black) errors. Our baseline DR6 birefringence band-powers and 1$\sigma$ statistical errors are shown in blue. The range of all other biases, plotted individually in Figure~\ref{fig:rot_biases}, is shown in orange. This figure demonstrates that while the range of possible global mean-field biases peaks above the other biases at $L<40$, they still are not quite significant compared to the 1$\sigma$ errors. Nevertheless, we report a separate conservative upper limit on the amplitude of cosmic birefringence where we choose $L_\mathrm{min} = 40.$}
    \label{fig:globmf}
\end{figure}
The ACT DR6 power spectrum analysis finds a global polarization angle offset of $0.2 \pm 0.1$~deg \citep{louis2025}. To quantify its effect, we generate sets of 100 simulations with uniform rotation angles of $\alpha = 0.2$, $0.3$, and $0.4$~deg, corresponding to the central value, $+1\sigma$, and $+2\sigma$ error of this measurement. These are otherwise identical to the transfer function simulations of Section~\ref{sec:tf}. We reconstruct anisotropic birefringence from each set and compute the resulting mean-fields. At $L < 40$, we find that the global-angle mean-field is comparable to or exceeds all other biases, including $N_L^{0,\alpha}$ (see Figure~\ref{fig:globmf}). We therefore split the analysis into two cases:

\begin{enumerate}
    \item Given that we represented the $2\sigma$ uncertainty in the global polarization angle in the test simulations, and quantified the spread in their mean-field biases for this birefringence analysis, we add the $1\sigma$ spread as a systematic error to our baseline $1\sigma$ errors (and the square of this $1\sigma$ spread to the diagonal of the birefringence covariance matrix). This enables us to use large reconstructed scales, and choose $L_\mathrm{min}^\alpha = 15$ to obtain the baseline result presented in Section \ref{sec:results}: $A_\mathrm{CB} = $\aCB, or a 95\% upper limit $A_\mathrm{CB} < $ \aCBUL\, where the mean-field bias contributes $0.01\sigma$ (see Figures \ref{fig:biref_bps} and \ref{fig:birefpost}). Arguably, with this method we could use all reconstructed scales, but because of the large uncertainty in the mean-field biases at $L < 15$, including the largest scales does not change the upper limit on $A_\mathrm{CB}$.
    
    \item We can choose not to incorporate any new error and just choose $L_\mathrm{min}^\alpha$ such that these global polarization angle mean-field biases are sub-dominant. Therefore, for our ``conservative'' analysis, we choose $L_\mathrm{min}^\alpha = 40$ which yields a 95\% upper limit of $A_\mathrm{CB} < 0.52$, as is also shown in Figure~\ref{fig:biref_bps}. This result demonstrates that the ability to reliably reconstruct large scales is crucial for improving constraints on a scale-invariant birefringence signal. 
\end{enumerate}

\subsection{Data-based Consistency Tests} \label{sec:const_tests}
Because we made several choices in these analyses, consistency tests are needed to ensure that the choices did not bias the final results. For each test, we use a different subset of the data or a different analysis choice to compute a new set of band-powers. We then compute differences between these band-powers that are expected to be consistent with null, given that there are no unmodeled contaminants. In each test, we also run the analysis on 600 - 792 simulations (depending on the dataset used) to obtain a difference covariance matrix and accurately estimate the errors on the difference. This simulation procedure enables us to build up a covariance matrix for each null test, in particular because for many of the null tests, common CMB sample variance between overlapping sets of data is removed. We summarize our results in Table \ref{tab:consistency_tests}, in Figure~\ref{fig:tau90150}, and in Appendix Figures~\ref{fig:masktest}, \ref{fig:ellrange_test}, and \ref{fig:rot90150}. In general, we find that all the cosmic birefringence consistency tests pass, but for screening we see some evidence for residual foreground contamination. We investigate this possibility with some of the tests detailed in this section.

\begin{table*}[]
\centering
\caption{Consistency tests for the screening ($\tau$) and cosmic birefringence ($\alpha$) analyses. The baseline result was subtracted from the band-powers from each test to construct null tests, except when we compare 90 GHz vs. 150 GHz band-powers. We report the $\chi^2$ and PTE for each null test.}
\label{tab:consistency_tests}
\begin{tabular}{lcccc}
\hline\hline
\textbf{Test} & $\bm{\tau}$: $\bm{\chi^2} (\mathrm{dof}=21)$ & $\bm{\tau}$\textbf{: (PTE)} & $\bm{\alpha}$: $\bm{\chi^2}(\mathrm{dof}=19)$ & $\bm{\alpha}$\textbf{: (PTE) }\\
\hline
Baseline Analysis Compared to Null  & 17.53 & 0.68 & 14.48 & 0.75 \\
\hline
$f_\mathrm{sky}$ Mask: 60\% (baseline) $\rightarrow$ 40\% & 14.8 & 0.83 & 10.2 & 0.95 \\
$f_\mathrm{sky}$ Mask: 60\% (baseline) $\rightarrow$ 70\%  & 32.7 & 0.05 & 20.6 & 0.36 \\
$\ell_\mathrm{min}: 600\ \mathrm{(baseline)} \rightarrow 800$  & 20.2 & 0.51 & 17.8 & 0.54 \\
$\ell_\mathrm{min}: 600\ \mathrm{(baseline)} \rightarrow 1000$  & 32.7 & 0.05 & 6.56 & 0.996 \\
$\ell_\mathrm{max}: 3000\ \mathrm{(baseline)} \rightarrow 2500$  & 18.5 & 0.62 & 13.5 & 0.81 \\
$\ell_\mathrm{max}: 3000\ \mathrm{(baseline)} \rightarrow 2000$   & 15.9 & 0.78 & 18.0 & 0.52 \\
Frequency Difference: 90 - 150 GHz  & 23.1 & 0.34 & 20.7 & 0.35 \\
CIB Deprojection: Deprojected - Baseline & 62.9 & 0.00 & -- & -- \\

\hline\hline
\end{tabular}
\end{table*}

First, we analyze the effect of the chosen mask (60\%) on the results compared to a more conservative (40\%) and more aggressive (70\%) mask. These percentages represent the amount of sky left in the data after masking the galaxy but before the ACT DR6 footprint mask is applied. For both screening and birefringence, these tests pass. The PTE for the 40\% mask in the birefringence analysis is slightly high, likely due to the increase in statistical error. The PTE for the 70\% mask in the screening analysis is fairly low at 0.05; however, we note that this is the ``aggressive'' mask, and therefore this test justifies the need to use the standard 60\% mask. 

\begin{figure*}
    \centering
    \includegraphics[width=0.49\linewidth]{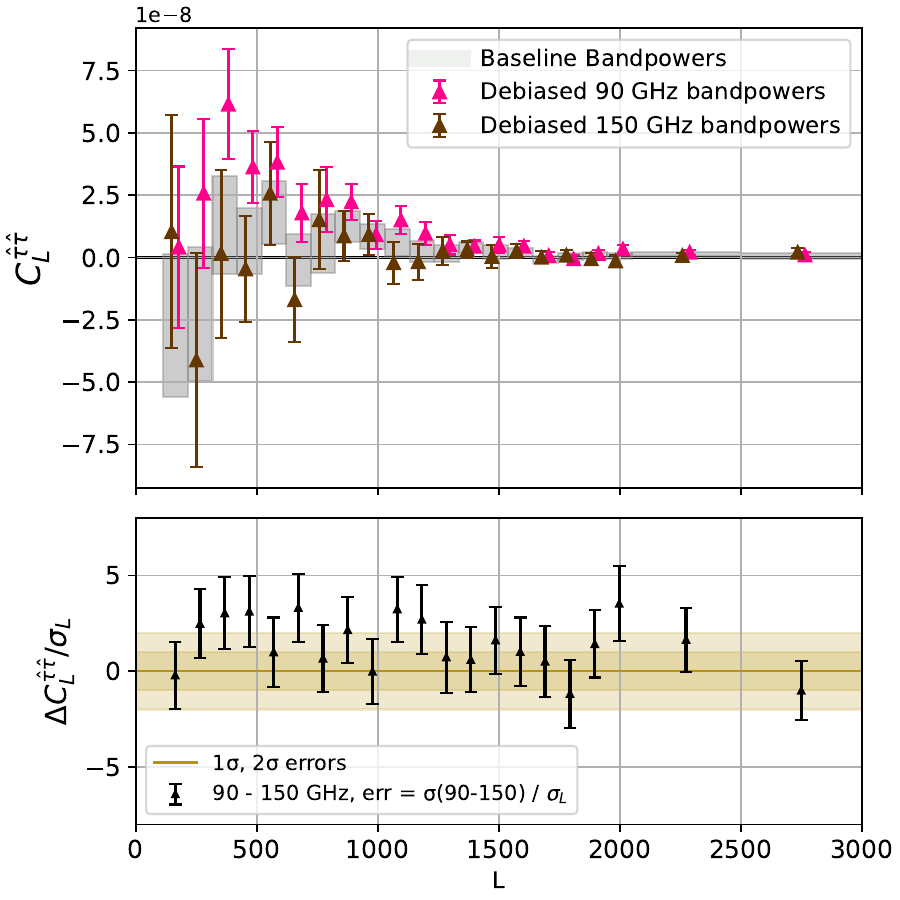}
    \includegraphics[width=0.49\linewidth]{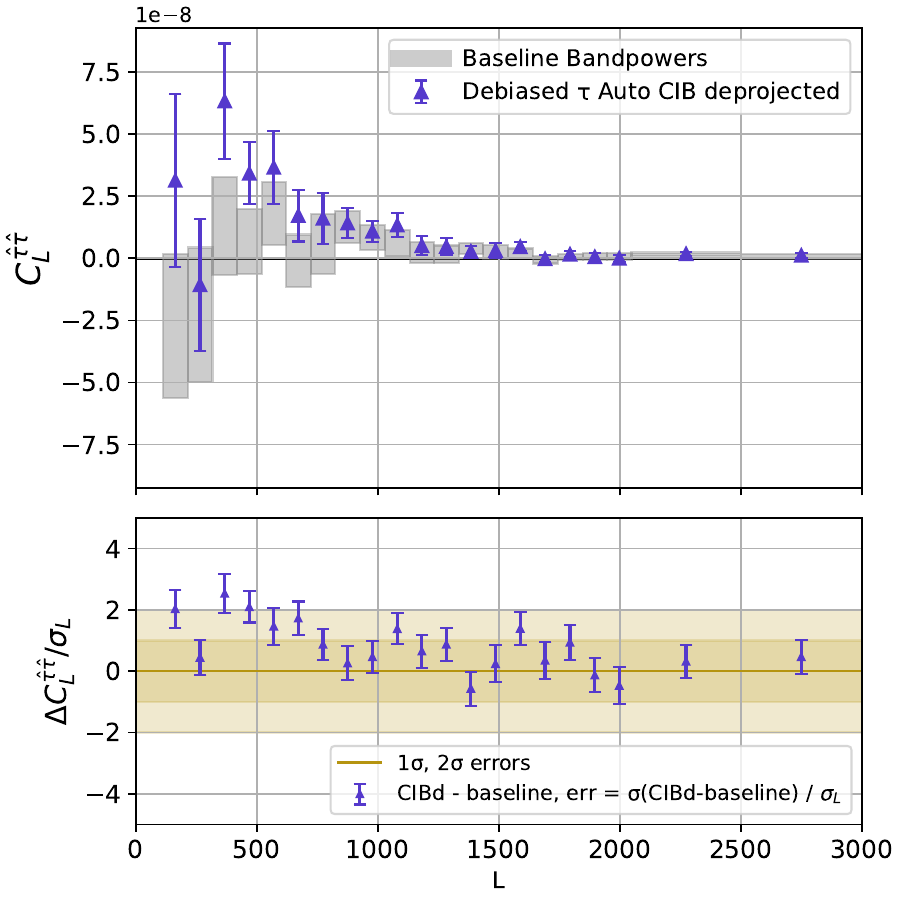}
    \caption{The two consistency tests from Table \ref{tab:consistency_tests} that are not consistent with null. \emph{Left: }Reconstructed $\tau$ band-powers from the 90 GHz-only (pink) and 150 GHz-only (brown) data compared to the coadd in gray boxes. Their difference null-test (black) is in the bottom panel with the 1 and 2$\sigma$ errors. Since the CMB has the same signal in 90 and 150 GHz, the difference is expected to be consistent with null unless foregrounds are strongly-measured. Indeed, this test is consistent with zero with a PTE of 0.34, but the test does not account for the constant bias in 90 GHz that is significant on its own compared to a null signal (PTE = 0.012). This excess is indicative of residual foregrounds in the 90 GHz data contaminating the reconstructed $\tau$ signal. Based on the investigation in Section \ref{subsec:fgs}, we conclude that this 90 GHz result is consistent with the \texttt{WEBSKY} result that 90 GHz foreground contamination will show up in the 90 GHz data but will not be as significant for the baseline 90 + 150 GHz results. \emph{Right: }The reconstructed $\tau$ band-powers for the CIB-deprojected analysis compared to baseline, represented in the same manner as the left panel. This null test failed with a PTE of zero (see the bottom row of Table \ref{tab:consistency_tests}). This is likely due to the CIB-deprojection process removing some lensing contamination, which in turn removes a majority of a negative foreground bias according to \texttt{WEBSKY}, and it therefore increases the amplitude of the measured spectrum significantly. Because of this result, we experimented with reducing $\ell_\mathrm{max}$ and report our findings in Table \ref{tab:cibdeproj_errs} and Figure~\ref{fig:lmax2000_tau}.}
    \label{fig:tau90150}
\end{figure*}

The next set of consistency checks were designed to test the robustness of the results to the CMB multipoles used in the analyses. We do two variations of each $\ell_\mathrm{min}$ and $\ell_\mathrm{max}$ compared to the baseline range, choosing $\ell_\mathrm{min}\in [800,1000]$ and separately $\ell_\mathrm{max}\in [2000,2500]$ for a total of four separate tests for screening and four for birefringence. We find that each of these null tests pass; however, we note that the birefringence PTE for the $\ell_\mathrm{min}= 1000$ case is  high: this is likely due to the loss of all modes that contribute to the SNR of this measurement, as demonstrated in Figure~\ref{fig:filters}. On the other hand, the screening PTE for this test is considerably low: this could be because this test isolates the foreground-contaminated CMB modes and removes many of the SNR-heavy modes from Figure~\ref{fig:filters}. We conclude that this test is consistent with our findings from Section~\ref{subsec:fgs}.

We also perform separate 90 and 150 GHz analyses and difference the two to check for significant changes that may be due to foregrounds. For both screening and birefringence, the $90 - 150$ GHz null tests pass. However, we note that for screening there was a significant bias in the 90 GHz analysis compared to \emph{null} with PTE = 0.012 (See the left panel of Figure~\ref{fig:tau90150}). This is consistent with our findings in Section~\ref{subsec:fgs}, based on the \texttt{WEBSKY} mocks, as well as the somewhat-low PTE $\ell_\mathrm{max}$ test from the previous paragraph. The remaining consistency test band powers are shown in the Appendix.

Finally, we perform a test that checks whether deprojecting a CIB template makes the measurement of anisotropic screening change significantly from our baseline result. This test involves using 400 simulations and four data splits that were generated from both the ACT DR6 and \textsl{Planck} data. Specifically, these simulations were built using a harmonic internal linear combination (hILC) method, and were first used in \citet{maccrann2024a} for the ACT DR6 lensing foreground analysis. They are informed by extra frequency channels that \textsl{Planck} has (here, 353 and 545 GHz), and these frequencies are combined in such a way that both minimizes the variance of the result and optionally deprojects a foreground using its assumed spectral energy distribution. Please see \citet{maccrann2024a} for more details about the construction of these simulations and data.

\begin{table}[]
    \centering
    \begin{tabular}{c|c|c|c}
       \hline
        \hline
      \hspace{-35pt}\makecell{\textbf{Analysis}\\ \textbf{Choices}}& $\mathbf{A_\tau}$ & \hspace{-35pt} \makecell{$\mathbf{1\sigma}$ \textbf{Statistical}\\ \textbf{Error}}& \hspace{-35pt} \makecell{\texttt{WEBSKY} \\ \textbf{Systematic}\\ \textbf{Error}} \\
    \hline
       \hspace{-35pt} \makecell{$\ell_\mathrm{max} = 3000$,\\ baseline} & $180$ & $\pm 157$ & $\pm^{213}_{186}$  \\
       \hline
       \hspace{-35pt} \makecell{$\ell_\mathrm{max} = 3000$,\\ CIB-deprojected} &  $512$ & $\pm 156$ & --- \\
       \hline
       \hspace{-35pt} \makecell{$\ell_\mathrm{max} = 2000$\\  } & $-67$ & $\pm 208$ &  $\pm^{35}_{56}$ \\
       \hline
       \hspace{-35pt} \makecell{$\ell_\mathrm{max} = 2000$,\\ CIB-deprojected} & $135$ & $\pm 219$ & --- \\
       \hline
    \end{tabular}
    \caption{A table with measurements of the amplitude of anisotropic screening, $A_\tau$, and the corresponding statistical and estimated systematic errors for various analysis choices. Specifically, $\ell_\mathrm{max}$ was varied from 3000 (baseline) to 2000, and the CIB-deprojected data was used in each $\ell_\mathrm{max}$ case. The values for $A_\tau$ and the statistical errors are obtained through the methods described in Section \ref{sec:tau_results}. The systematic error range is estimated from the full range of \texttt{WEBSKY} foreground bias terms calculated in Section \ref{subsec:fgs}. This table demonstrates that deprojecting the CIB biases $A_\tau$ high in both $\ell_\mathrm{max}$ cases, but does so outside of the $1\sigma$ statistical range for the baseline analysis and well within the same range for $\ell_\mathrm{max} = 2000$. We therefore report $A_\tau <$\ \atauconsUL\ with $\ell_\mathrm{max} = 2000$ as an alternative, more conservative result.}
    \label{tab:cibdeproj_errs}
\end{table}

We find that when deprojecting the CIB, and reconstructing screening using $\ell_\mathrm{CMB,max} = 3000$, there is a significant increase in power compared to baseline (see the final row in Table \ref{tab:consistency_tests}  with $\chi^2 = 62.9$ for 21 degrees of freedom (PTE $< 10^{-5}$) and the right panel of Figure~\ref{fig:tau90150}). This is counterintuitive at first; one would expect that the deprojection of a foreground would reduce the bias and make the band-powers more consistent with zero. However, we interpret this increase as potentially due to the following: since the primary bispectrum from lensing and other foregrounds is expected to have a \emph{negative} bias on $C_L^{\hat{\tau}\hat{\tau}}$ according to \texttt{WEBSKY}, removing the CIB potentially removes a portion of this negative bias, leaving behind other positive foreground biases. (See Figure~\ref{fig:websky} for the \texttt{WEBSKY}-based estimates of the foreground bias terms for different analysis choices.) This therefore explains the increase in the screening band-powers when the CIB is deprojected. To confirm this result, we run the CIB deprojected analysis with $\ell_\mathrm{max} = 2000$ and indeed find that the result is consistent with the standard $\ell_\mathrm{max} = 2000$ analysis  (PTE = 0.53).

\begin{figure}
    \centering
    \includegraphics[width=\linewidth]{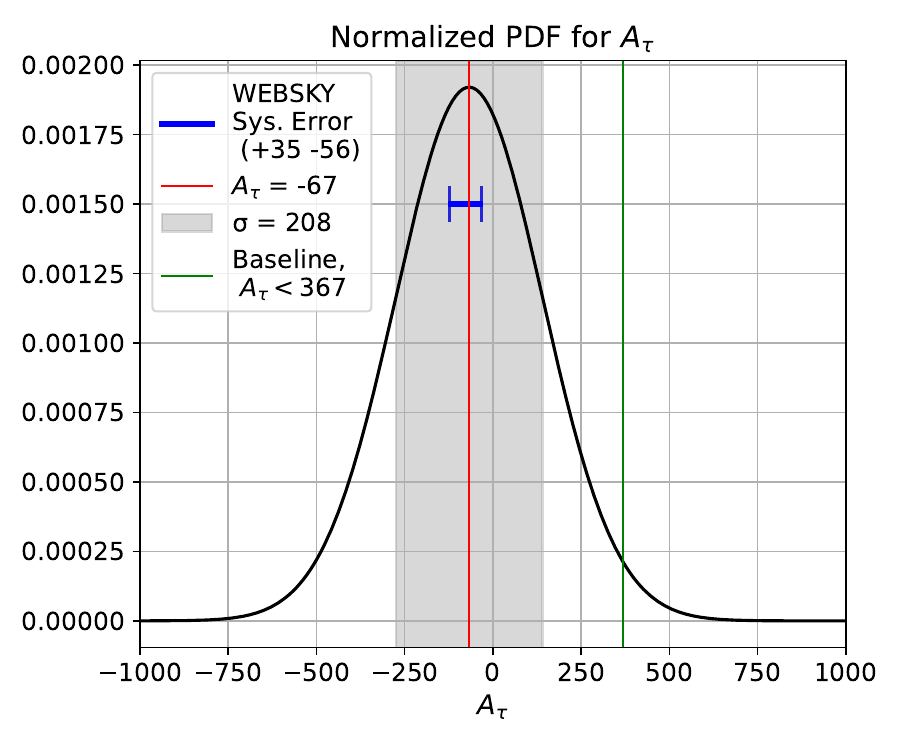}
    \caption{The posterior on $A_\tau$ when $\ell_\mathrm{max} = 2000$ instead of 3000 (see Figure~\ref{fig:taupost}.)
    This analysis option reduces the systematic foreground bias expected from the \texttt{WEBSKY} simulation, and is therefore more conservative than our baseline result. We again make no detection, and we place a 95\% confident upper limit of $A_\tau \lesssim$\ \atauconsUL. This upper limit is more stringent than baseline $\ell_\mathrm{max}=3000$ because this central value scattered low.}
    \label{fig:lmax2000_tau}
\end{figure}

This CIB-deprojected result, along with the excess power in the 90 GHz data, point to strong systematic errors from foreground emission that may be significant at ACT DR6 sensitivities. These findings warranted deeper exploration of our maximum CMB multipole choice in the \texttt{WEBSKY} foreground residual calculations, and we show the results of this investigation in the right panel of Figure~\ref{fig:tau90150}. Decreasing $\ell_\mathrm{max}$ from 3000 to 2000 reduces the \texttt{WEBSKY} systematic range while increasing the statistical error, so the systematic bias on $A_\tau$ becomes less significant relative to $\sigma(A_\tau)$. We therefore report a conservative $\ell_\mathrm{max} = 2000$ result for $A_\tau$, shown in the third row of Table \ref{tab:cibdeproj_errs} and in Figure~\ref{fig:lmax2000_tau}. We again find a signal consistent with zero (PTE = 0.98, see the band-powers in Figure~\ref{fig:ellrange_test}), and obtain an upper limit of $A_\tau <$\ \atauconsUL\ at 95\% confidence. This result is more stringent than our baseline one because the central value scattered low, but in either case we do not detect any signal.

\section{Discussion}\label{sec:disc}

In the following section we interpret our findings and discuss prospects for studies with current and future data.

\subsection{Anisotropic Screening Discussion}\label{subsec:taufuture}
Consistent with previous upper limits from \textsl{WMAP}, \textsl{Planck}, and BICEP/Keck, the ACT DR6 upper limit on $A_\tau$ does not yet reach the amplitude range predicted by current reionization models. Here, we discuss the main limiting factors and the prospects for improvements with upcoming data.

In this work we find that screening analyses have a level of contamination from extragalactic foregrounds that is potentially on the order of the current statistical uncertainty (see Section \ref{subsec:fgs}). Given our significant uncertainties on the non-Gaussianity of the much brighter CIB and tSZ effects, with future high-resolution data it will be necessary to direct efforts to foreground removal or avoidance before screening can be significantly measured. One way to avoid the higher-point functions of CMB temperature foregrounds in a screening analysis is to incorporate the polarization-only estimator, $\hat{\tau}(EB)$. Polarization foregrounds are much weaker than those in temperature, and a polarization-only analysis of anisotropic screening would yield a much lower foreground bias. This method was not used here because the ACT DR6 polarization noise, as seen in our corresponding birefringence analysis, is relatively large and a polarization-based reconstruction would not improve current upper limits or be competitive with our temperature-based results. SO and South Pole Observatory (SPO) data, however, will have improved polarization noise, and therefore will make more robust measurements of screening. 

Another important tool in foreground studies is multi-frequency data. The \textsl{Planck} data has revolutionized CMB foreground studies, and combining it with ground-based data such as ACT will not only improve constraining power but will also reduce the foreground contamination to the result. Future ground-based, multi-frequency data from SO, the South Pole Observatory (SPO), and the Cornell-Caltech Atacama Telescope \citep[CCAT,][]{ccat2018}, will also significantly improve the outcomes of foreground separation analyses thanks to their higher resolution and more frequency coverage than current surveys.

A better theoretical understanding of foreground biases to screening estimators would also be valuable. Here, we have only studied the \texttt{WEBSKY} model, and only for the sensitivities and frequency combinations of the ACT DR6 data. Future work could involve evaluating other recent models of the non-Gaussian mm-wave sky on small scales, such as those encoded in the \texttt{Agora} \citep{Omori2024} or \texttt{FLAMINGO} \citep{yang2025} mocks, in the context of upcoming datasets.

Finally, based on Figure~\ref{fig:filters}, the important CMB scales for reconstructing screening are well-measured by ACT DR6.  The integrand is nonetheless larger at 
 $\ell_\mathrm{min} = 600$ than at
$\ell_\mathrm{max} = 3000$, so the modes lost on large scales more than those lost at small scales. This means it is slightly more favorable to measure the largest CMB scales for this type of measurement, something that space-based satellites such as \textsl{Planck} do best. Therefore, a future screening analysis with ACT+\textsl{Planck} data will yield tighter constraints. 

\subsubsection{Implications for Reionization}\label{subsec:reionization_future}

\begin{figure*}
    \centering
    \includegraphics[width=0.45\textwidth]{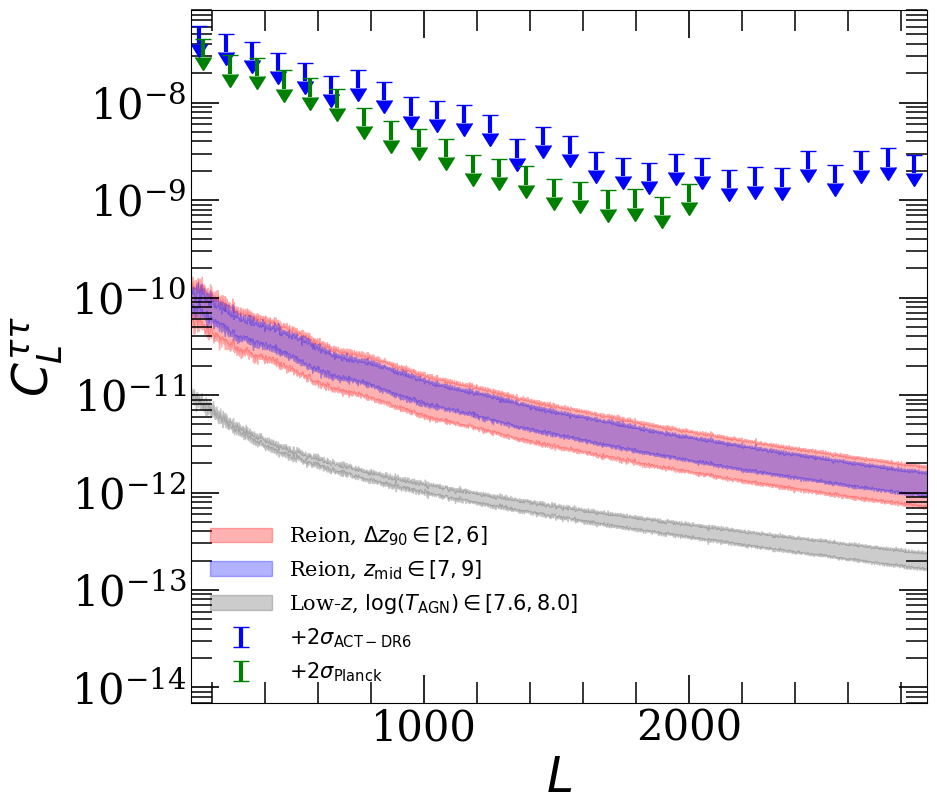}
    \includegraphics[width=0.54\textwidth]{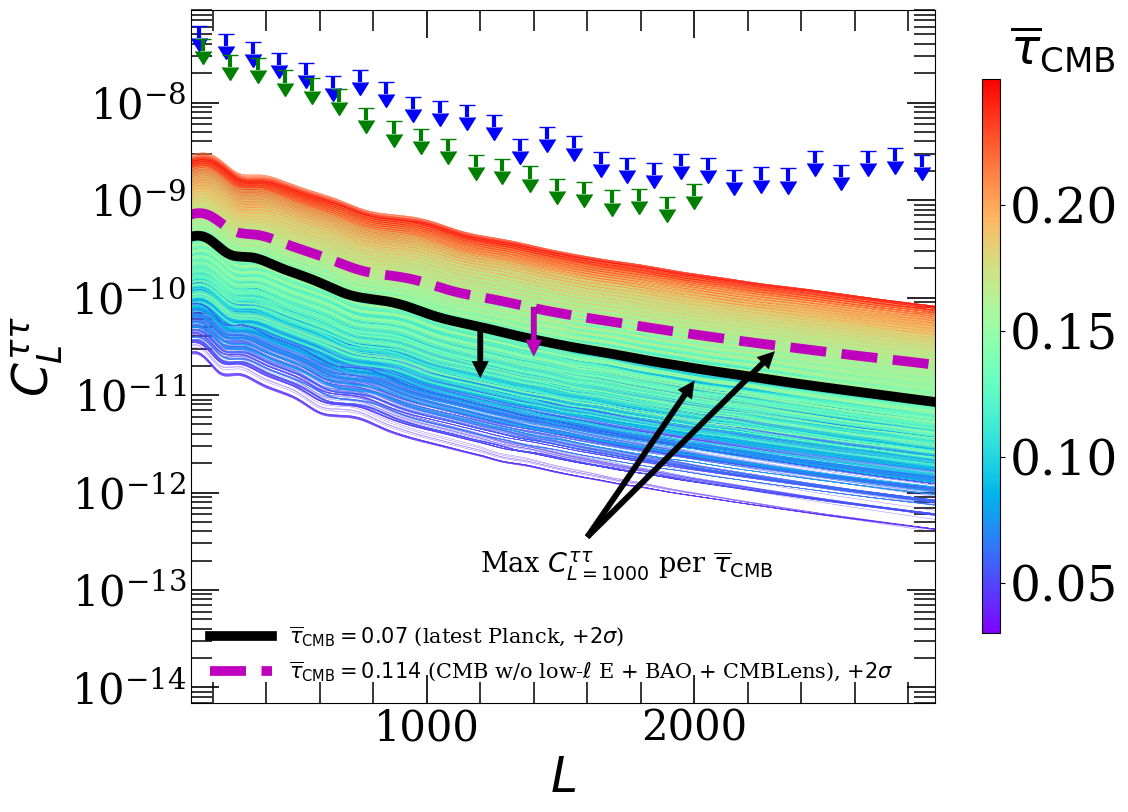}
    \caption{Assessment of how much sensitivities on the $C_{L}^{\tau\tau}$ measurement need to improve before theoretically interesting limits can be placed on the signal. \emph{Left:} $2\sigma$ sensitivities for \textsl{Planck} (green) and ACT (blue) alongside the range of fiducial theoretical models shown in Figure~\ref{fig:tau_bps}. The red (blue) bands show ranges of $\Delta z$ ($z_{\rm mid}$). \emph{Right:} The same sensitivities shown against a much wider range of theory predictions, color-coded their $\overline{\tau}_{\rm CMB}$. The thick lines show the models we found with the highest $C_{L=1000}^{\tau\tau}$ for a given upper limit on $\overline{\tau}_{\rm CMB}$, given in the legend. See text for further details.}
    \label{fig:ACT_limits_plot}
\end{figure*}

It is instructive to consider how far current measurements are, in terms of sensitivity, from being able to rule out (or confirm) models of theoretical relevance. In the left panel of Figure~\ref{fig:ACT_limits_plot}, we show our $2\sigma$ sensitives alongside those found with \textsl{Planck}\ \citep{toshiyatau2018}, compared to the same \texttt{AMBER} models shown in Figure~\ref{fig:tau_bps} (only here not multiplied by  a factor of 100). We see that models reasonably close to nominal theoretical expectations are more than $2$ orders of magnitude below our current sensitivities at all $L$ values. This pessimistic result indicates that current experiments are far from making a detection if reionization follows standard expectations.

We find a slightly more optimistic picture for current data if we instead ask the question, ``how far away are we from ruling out an interesting reionization model?''  To answer this question, we broaden our search of model space to include scenarios with much higher mean optical depths $\overline{\tau}_{\rm CMB}$ than that of our fiducial model. Specifically, we run \texttt{AMBER} models across a regular grid in the reionization history parameters, spanning redshift midpoint $z_{\rm mid} \in [6.5,15.5]$, duration $\Delta z \in [2,31]$, and asymmetry $A_z \in [1,20]$, with the step size along each parameter being unity. Thus, we survey a total of $10 \times 30 \times 20 = 6000$ models, with $\overline{\tau}_{\rm CMB}$ ranging from $\approx 0.04$ to $0.22$ (see the right panel of Figure~\ref{fig:ACT_limits_plot}). We then consider which values of $\overline{\tau}_{\rm CMB}$ are allowed by recent measurements in the literature. In particular, the latest $2\sigma$ upper limit on $\overline{\tau}_{\rm CMB}$ from \textsl{Planck} is $0.07$~\citep{Tristram2024}, while the similar upper limit inferred by~\citet{Sailer2025} (which uses recent DESI-DR2 baryon acoustic oscillations (BAO), primary CMB, and CMB lensing data, but excludes low-$L$-EE) is $0.114$. For each of these $\overline{\tau}_{\rm CMB}$ values, we identify which of our models has the highest $C_{L=1000}^{\tau\tau}$, conditioned on this value of $\overline{\tau}_{\rm CMB}$. 

These are shown by the thick lines, with downward arrows indicating that they bound $C_L^{\tau\tau}$ from above, within our model grid, for a given upper limit on $\overline{\tau}_{\rm CMB}$.

We see that surveying a wider parameter space reveals models with much higher $C_{L}^{\tau\tau}$ than our fiducial models that are allowed by current $\overline{\tau}_{\rm CMB}$ measurements. In particular, these are models with large $\Delta z$ (and thus high $C_{L}^{\tau\tau}$, but also large $A_z$, such that they have low $z_{\rm mid}$ (and thus low $\overline{\tau}_{\rm CMB}$). We see that the model with the highest $C_{L}^{\tau\tau}$ that is allowed by the~\citet{Sailer2025} limit on $\overline{\tau}_{\rm CMB}$ has nearly an order of magnitude higher signal than our fiducial scenarios, and only about an order of magnitude away from \textsl{Planck} upper limits. This suggests that an order of magnitude improvement in current sensitivities could yield theoretically interesting upper limits on the $C_{L}^{\tau\tau}$ signal, provided foregrounds can be mitigated at a similar level. Sensitivities from upcoming CMB surveys will significantly reduce these upper limits and may even make detections of this signal \citep[see, e.g., ][]{dvorkinsmith2009, Bianchini2023, Jain2024}. These future screening studies will be especially helped by deeper CMB polarization data, whose foregrounds are much less problematic than those in the CMB temperature signal.

\subsection{Anisotropic Cosmic Birefringence Discussion}\label{subsec:rotfuture}
The ACT DR6 data show no evidence for anisotropic cosmic birefringence ($A_\mathrm{CB} =$\ \aCB), consistent with all other current measurements, with which the constraining power is comparable. Our upper limit of $A_\mathrm{CB} <$\ \aCBUL\ can be translated to an upper limit on the Chern-Simons coupling term $g_{a \gamma} < $\ \gUL. The persistent hint of a nonzero \emph{isotropic} rotation angle \citep[e.g.,][]{minamikombirefnonzero, diegopalazuelos2022} motivates continued searches for an anisotropic component. Depending on the underlying mechanism, models that produce a uniform rotation can also generate spatially-varying fluctuations at levels within reach of near-term experiments \citep[e.g.,][]{pogosian2019}. Current upper limits on $A_\mathrm{CB}$, including the one presented here, are beginning to constrain the parameter space of axion-like models but do not yet rule out the favored mass ranges.

The main systematic to cosmic birefringence in this analysis is the mean-field bias generated from a global polarization angle (see Figure~\ref{fig:globmf}). Its amplitude and uncertainty are measurable via independent methods with primary CMB power spectrum data for a given instrument. By adding the uncertainty to our baseline analysis uncertainties via a mean-field bias, we were able to push $L_\mathrm{min}^\alpha$ down to 15. This significantly strengthens the upper limit inferred by $C_L^{\hat{\alpha}\hat{\alpha}}$  compared to the $L_\mathrm{min}^\alpha = 40$ case, from $\sigma(A_\mathrm{CB}) = $ 0.2 to 0.06 for a scale invariant signal, which peaks on large scales. 

Even more than screening, anisotropic cosmic birefringence is best measured using low-noise, large-scale CMB polarization data (see Figure~\ref{fig:filters}). \textsl{Planck} \citep{Bortolami_2022} provide the tightest upper limits on $A_\mathrm{CB}$, closely followed by BICEP/Keck \citep{Ade2023}, demonstrating that large CMB scales are necessary for constraining anisotropic cosmic birefringence. However, ACT and SPT are competitive in constraining power. Consequently, the most promising analyses with current data likely involve combining large scale and high resolution/low noise CMB data. Future ground and space-based surveys, such as SO and a potential future CMB space mission, \textsl{PICO} \citep{pico2019}, will significantly improve current upper limits on $A_\mathrm{CB}$, and could additionally make strong detections of the isotropic birefringence angle if it is consistent with current hints \citep{pogosian2019,litebird_isobiref_forecast_2025}.

We note that this ACT DR6 analysis reports a higher upper limit than the previous ACT DR4 analysis in \citet{namikawa2020}. This is due to the different datasets: DR4 was a smaller but deeper patch of the CMB, allowing for reduced polarization noise and the use of larger CMB scales than used in this work. Furthermore, we use the cross-split estimator in this work to guard against any noise mis-modelling that may have occurred, which is more likely in a larger area survey. As mentioned in Section~\ref{subsec:biref_results}, in ensuring an un-biased result to noise anisotropy this estimator induces slightly larger variance. The ACT DR4 analysis used the standard coadd estimator.

\section{Conclusions} \label{sec:conc}
In this paper, we used the ACT DR6 data to place limits on the anisotropic screening and birefringence effects. We expanded the pipeline developed for CMB lensing analyses to create reconstructed, noisy maps of these effects. We then computed the power spectra of these maps, and to obtain unbiased measurements we studied biases unique to screening and birefringence, mitigated all known biases including a full accounting of the impacts of CMB lensing on our estimators, and performed several consistency checks. We found our band-powers to be consistent with a null signal in both cases. We then reported 95\% upper limits on both signals, finding $A_\tau < $\ \atauUL\ and $A_\mathrm{CB} <$\ \aCBUL\ ($g_{a \gamma} <$ \gUL) in our baseline analyses. Both upper limits are consistent and comparable with constraints from other recent CMB surveys on these signals. However, the methods we developed here will place tighter constraints on these effects with more sensitive and specialized datasets.

This work represents the first analysis of the power spectrum of anisotropic screening using the small scales accessible with ground-based CMB surveys. As a result of using these smaller scales, we discovered that this type of analysis is potentially sensitive to residual foreground components; there is excess power in the 90 GHz reconstruction compared to the 150 GHz and coadd reconstructions. To investigate this, we studied simulated \texttt{WEBSKY} residual foreground maps to assess the components that are likely plaguing the screening reconstructions from 90, 150, and 90 + 150 GHz and we found a similar excess at 90 GHz. With thorough testing, we determined that the systematic error on screening from temperature foregrounds may be comparable to the statistical error from ACT DR6. Removing the CIB from the data (using \textsl{Planck}) skews $A_\tau$ high, likely due to the elimination of a negative foreground term. Conservatively reducing the $\ell_\mathrm{max}$ used in the reconstruction to 2000 significantly reduces the range of systematic foreground error as estimated by \texttt{WEBSKY}, yielding a 95\% upper limit on $A_\tau$ of \atauconsUL. In light of these findings, future screening studies will benefit from polarization-based estimators and improved foreground-mitigation techniques as upcoming surveys achieve improved polarization sensitivity and higher resolution multi-frequency data. Upcoming CMB surveys are expected to achieve the first detections of this signal for realistic reionization scenarios \citep[e.g.,][]{roy2018,Bianchini2023,Jain2024}.

For birefringence, we found no hints of uncharacterized systematics that dominate the DR6 measurement. We presented two sets of limits on the amplitude of the birefringence spectrum, based on differing treatments of our leading systematic effect, which arises from a non-zero global rotation angle in the CMB polarization data. These were $A_\mathrm{CB} <$ \aCBUL\ for our baseline treatment, and  $A_\mathrm{CB} < 0.52$ for our conservative analysis. Our baseline upper limit corresponds to the Chern-Simons coupling term $g_{a \gamma} <\ $\gUL. We determined that this measurement would significantly improve from using the same methods on combinations of current ground and space-based data, as well as lower-noise polarization data from both regimes.

With upcoming CMB surveys, such as SO and SPO, we will uncover brand new secondary CMB anisotropies that will tell us about the Universe. In this paper, we searched for two of them, anisotropic screening and birefringence, using the latest data and optimal methods. Though they remain undetected, the outlook for detecting these signals with data from planned CMB surveys is promising. It will yield brand new information about reionization, baryonic effects, and beyond-standard-model physics.

\section*{Acknowledgments}
Support for ACT was through the U.S.~National Science Foundation through awards AST-0408698, AST-0965625, and AST-1440226 for the ACT project, as well as awards PHY-0355328, PHY-0855887 and PHY-1214379. Funding was also provided by Princeton University, the University of Pennsylvania, and a Canada Foundation for Innovation (CFI) award to UBC. ACT operated in the Parque Astron\'omico Atacama in northern Chile under the auspices of the Agencia Nacional de Investigaci\'on y Desarrollo (ANID). The development of multichroic detectors and lenses was supported by NASA grants NNX13AE56G and NNX14AB58G. Detector research at NIST was supported by the NIST Innovations in Measurement Science program. Computing for ACT was performed using the Princeton Research Computing resources at Princeton University, the National Energy Research Scientific Computing Center (NERSC), and the Niagara supercomputer at the SciNet HPC Consortium. SciNet is funded by the CFI under the auspices of Compute Canada, the Government of Ontario, the Ontario Research Fund–Research Excellence, and the University of Toronto. We thank the Republic of Chile for hosting ACT in the northern Atacama, and the local indigenous Licanantay communities whom we follow in observing and learning from the night sky.

DK and AvE were supported by NASA grant 80NSSC24K0665.  DK and AvE were additionally supported by NASA grants 80NSSC23K0747 and 80NSSC23K0464, and NSF AAG grant 588167.   AvE thanks the Kavli Institute for Cosmology Cambridge for their hospitality during his stay as a Kavli Medium-Term Visitor,  during which part of this work was performed.    IAC acknowledges support from Fundaci\'on Mauricio y Carlota Botton and the Cambridge International Trust. CC acknowledges support from the Beus Center for Cosmic Foundations at Arizona State University. CS acknowledges support from the Agencia Nacional de Investigaci\'on y Desarrollo (ANID) through Basal project FB210003. NS acknowledges support from DOE award number DE-SC0025309. MM and AvE acknowledge support from NSF grant AST-2307727.
% \newpage
\bibliography{BIB}

\appendix
Below are the results of the band-power null tests from Table \ref{tab:consistency_tests} that passed with PTEs $\leq 1$ or $\geq 0.05$. These tests are introduced in Section \ref{sec:const_tests}. Figure~\ref{fig:covmats} shows the correlation matrices for the screening and birefringence baseline analyses in this work, briefly discussed in Section \ref{sec:covariance}.

\begin{figure}[h!]
    \centering
    \includegraphics[width=0.53\linewidth]{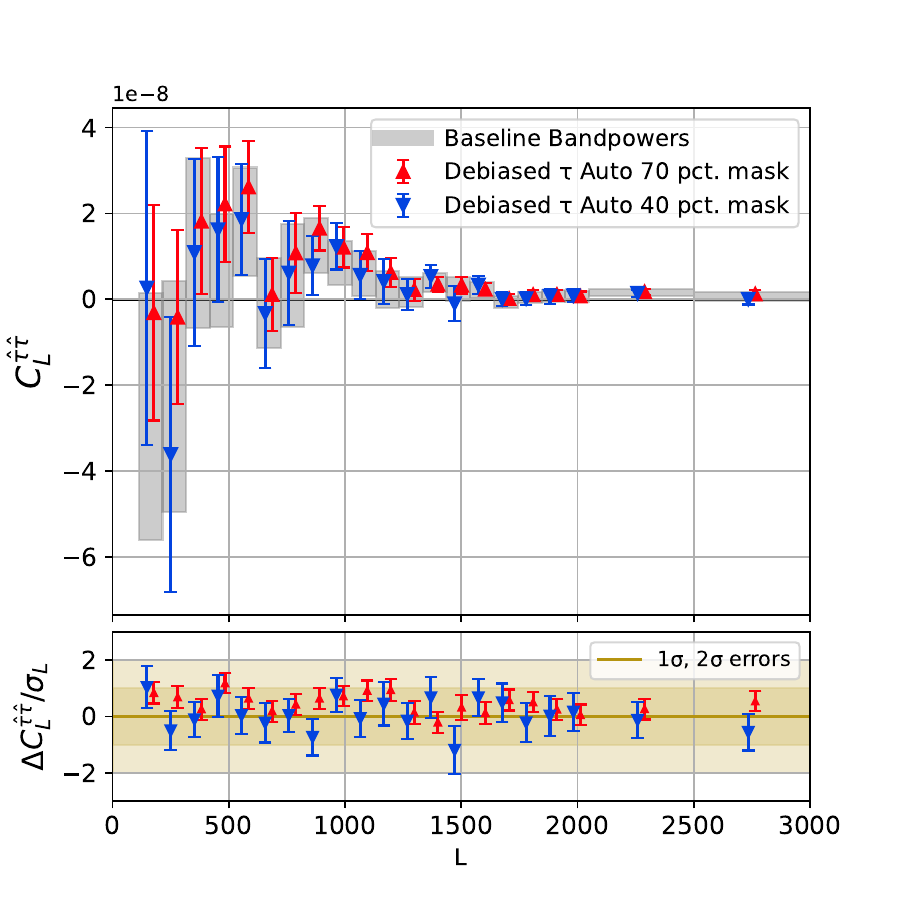}
    \hspace{-20pt}
    \includegraphics[width=0.49\linewidth]{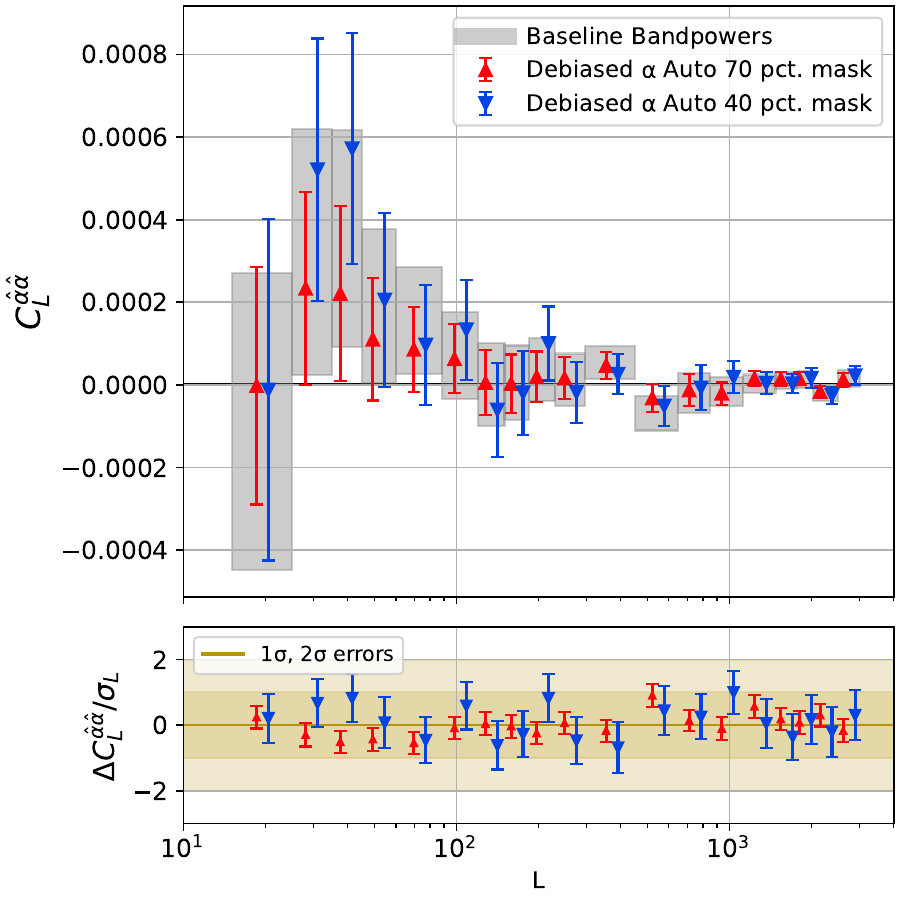}
    \caption{Test of robustness to mask variation on the anisotropic screening (left) and cosmic birefringence (right) band-powers. Red band-powers represent those reconstructed using a more aggressive galaxy mask, increasing the sky fraction from the baseline analysis. The blue band-powers are the result of a more restrictive mask, reducing the sky fraction from the baseline analysis. In both cases, the difference from the individual test to the baseline band-powers (represented by the gray boxes) is consistent with null. For the screening 70\% mask test, the PTE is 0.05, meaning it marginally passes. This is not worrisome because we do not use the 70\% mask in the actual analysis due to possible Galactic foreground sensitivity. Therefore, it confirms that we chose the proper mask for our baseline analysis. }
    \label{fig:masktest}
\end{figure}

\begin{figure}
    \centering
    \includegraphics[width=0.45\linewidth]{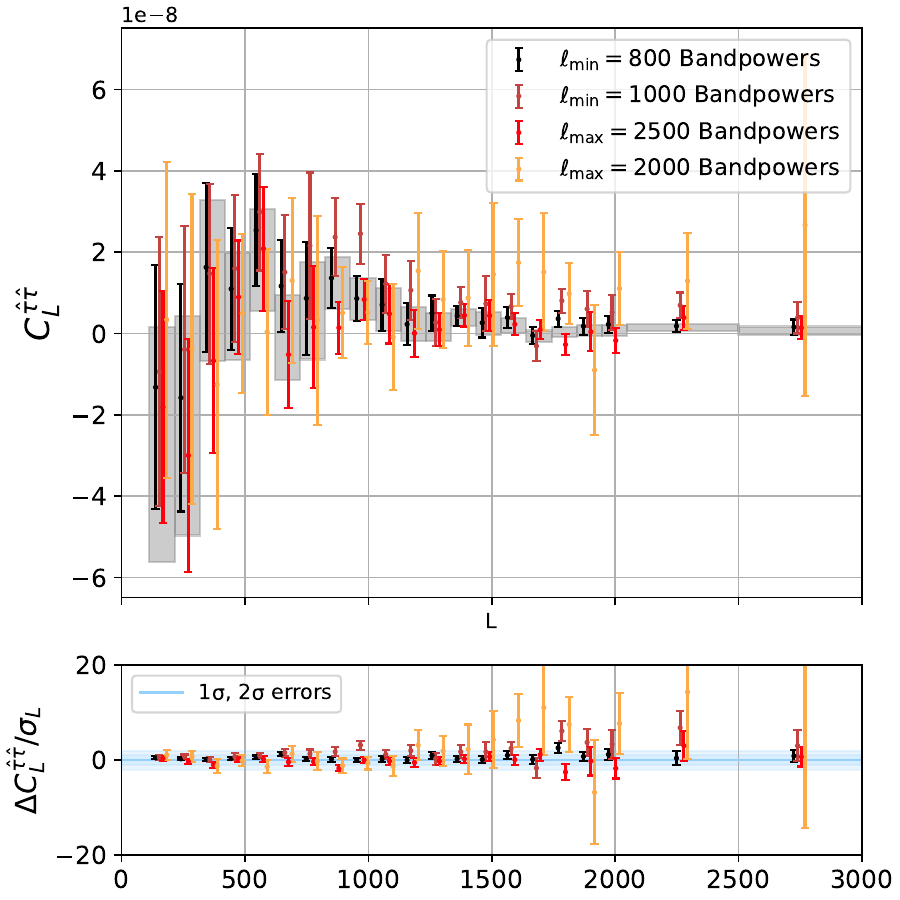}
    \includegraphics[width=0.45\linewidth]{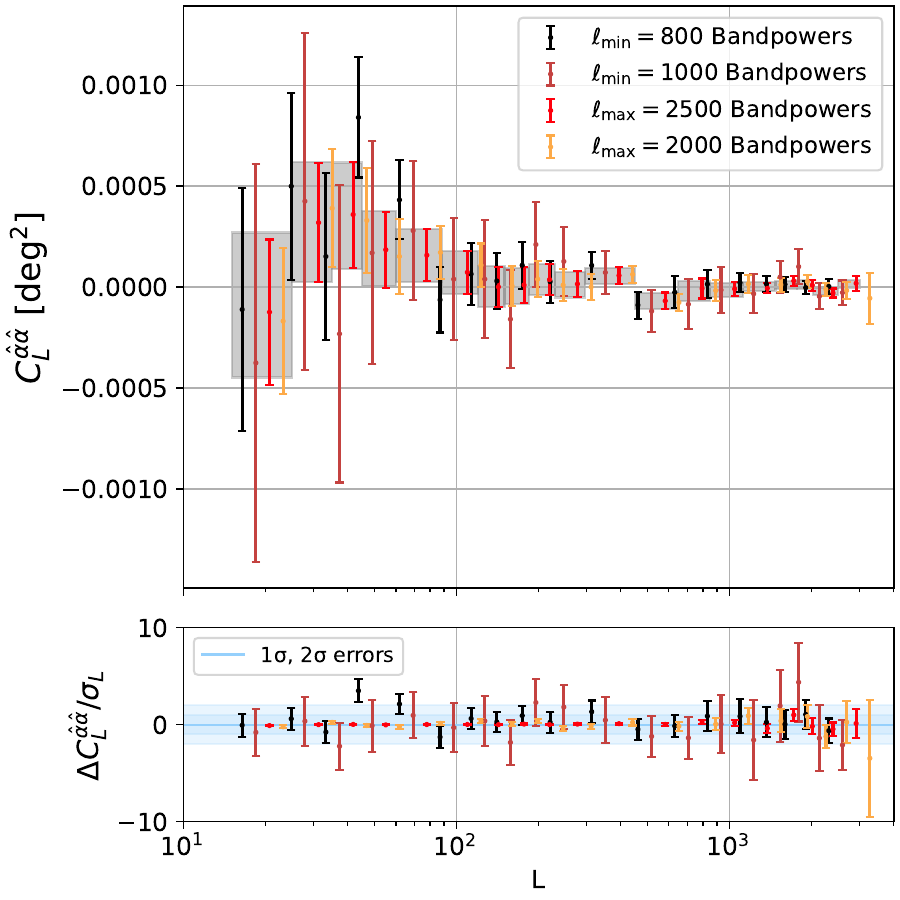}
    \caption{Band-powers and null tests using different ranges of CMB multipoles compared to the baseline range of $600 < \ell < 3000$ for both screening (\emph{left panel}) and birefringence (\emph{right panel}). The PTEs are listed in Table \ref{tab:consistency_tests}, and most are un-notable except for the screening $\ell_\mathrm{min} = 1000$ case (PTE = 0.05), which is insensitive to important scales for measuring screening and is sensitive to CMB scales where foregrounds begin to matter. The fact that foregrounds may be affecting this particular null test is consistent with our findings in Section \ref{subsec:fgs}.}
    \label{fig:ellrange_test}
\end{figure}

\begin{figure}[h]
    \centering
    \includegraphics[width=0.5\linewidth]{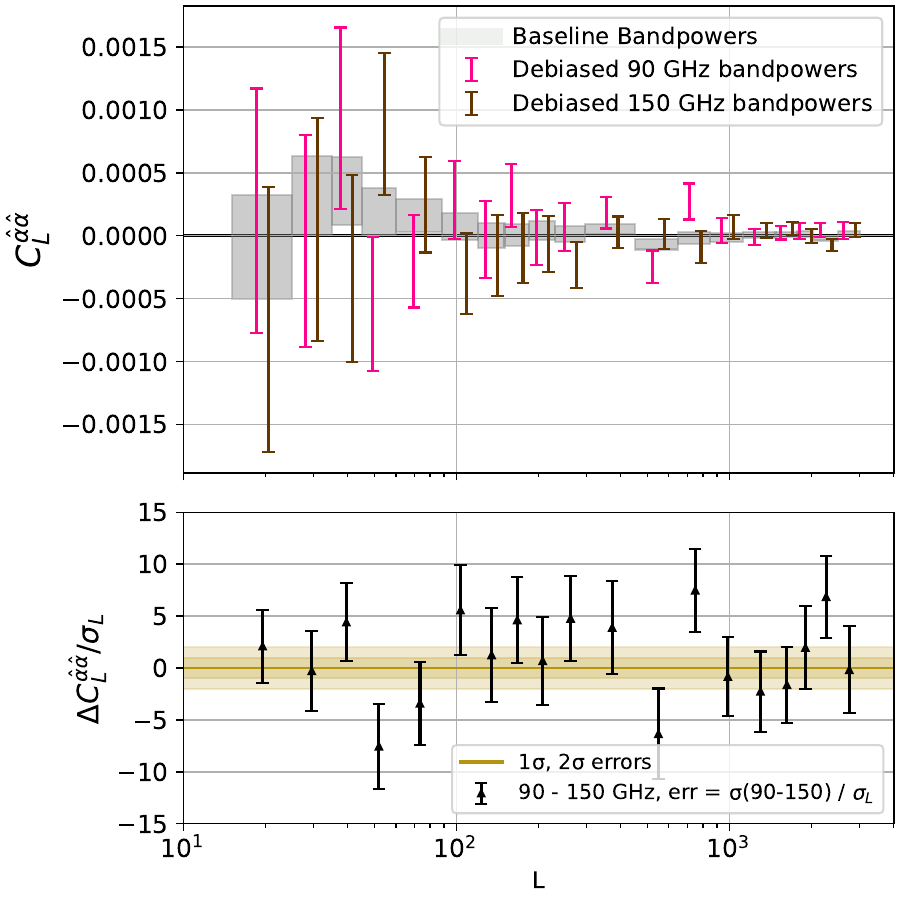}
    \caption{90 GHz (pink) and 150 GHz (brown) band-powers (top panel) and their difference null test (bottom panel) for the cosmic birefringence analysis. This test was consistent with null, with a PTE of 0.35. Since neither frequency shows a significant bias, this indicates that frequency-dependent foreground contamination in the ACT DR6 data is not significant enough to affect the results.}
    \label{fig:rot90150}
\end{figure}

\begin{figure}
    \centering
    \includegraphics[width=0.49\linewidth]{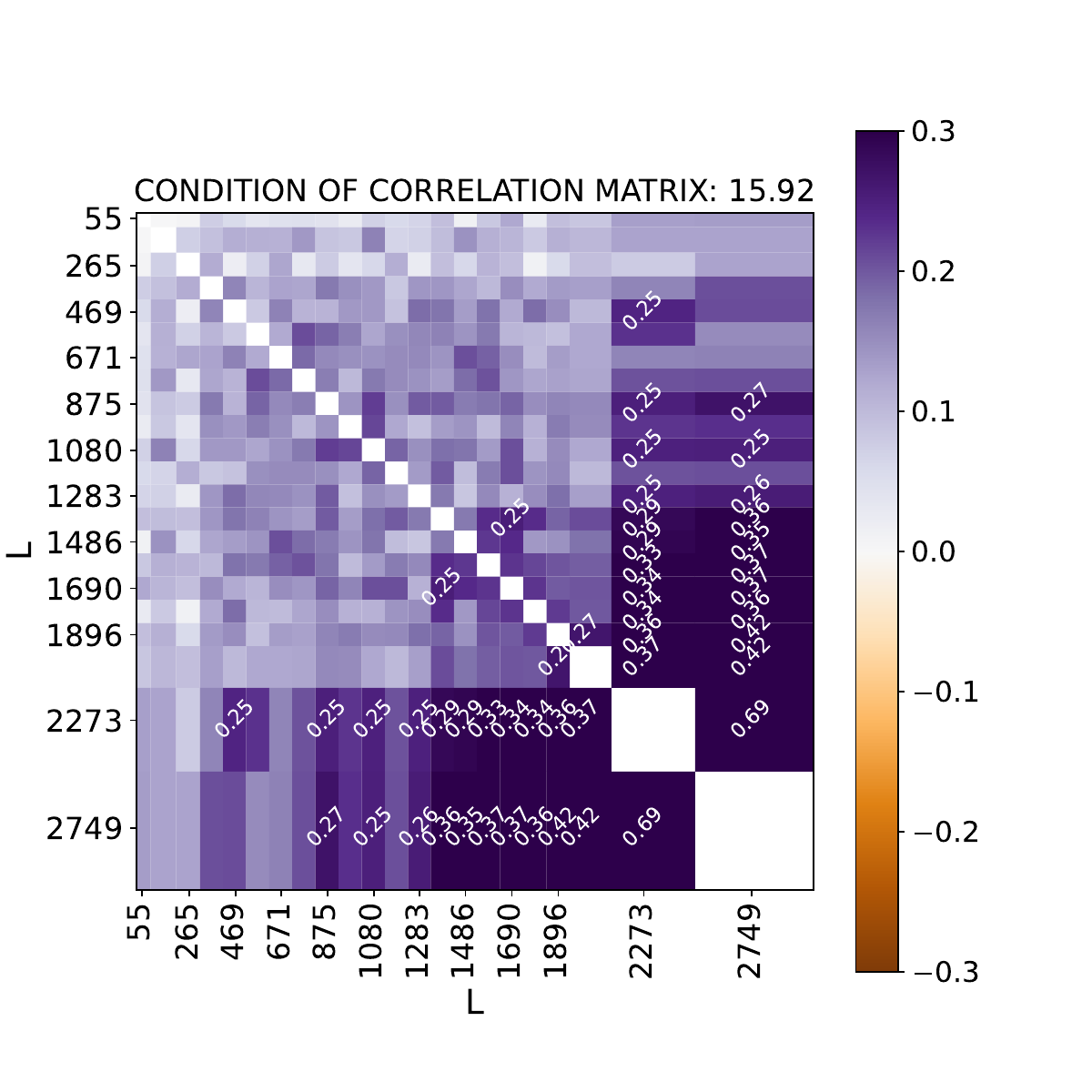}
    \includegraphics[width=0.49\linewidth]{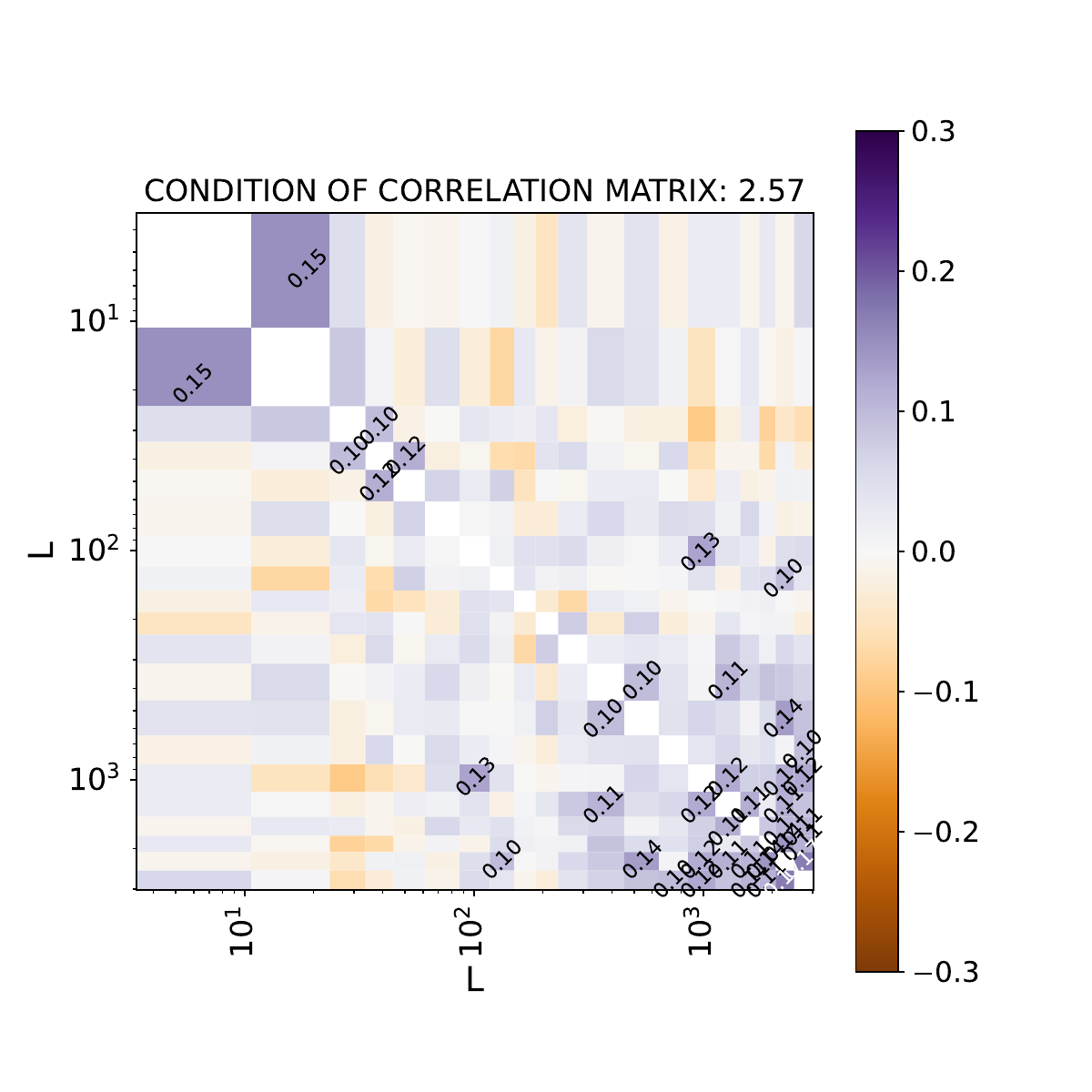}
    \caption{The correlation matrices for the baseline analysis in this work. \emph{Left:} The anisotropic screening correlation matrix, showing the large covariance at small reconstructed scales. \emph{Right:} The anisotropic birefringence correlation matrix, in log-log space due to our binning choice, demonstrating a lack of significant covariance at most reconstructed scales. We built each of these from reconstructions of 792 simulations. We show their respective conditions above each, and note that both were within acceptable range to deem them invertible for the error bar and likelihood analyses.}
    \label{fig:covmats}
\end{figure}
\end{document}